\documentclass[aps,prb,twocolumn,floatfix,nofootinbib,superscriptaddress,longbibliography,nobibnotes]{revtex4-2}

\usepackage{amsmath}
\usepackage{amssymb}
\usepackage{mathtools}

\usepackage[T1]{fontenc}
\usepackage{amsfonts}
\usepackage{newtxtext}
\usepackage[varvw]{newtxmath}
\usepackage{dsfont}                 
\usepackage{bbold}                  
\usepackage[normalem]{ulem}         

\usepackage{array}
\usepackage{dcolumn}                
\usepackage{siunitx}

\usepackage{enumerate}
\usepackage[shortlabels]{enumitem}

\usepackage[usenames,dvipsnames,table]{xcolor}
\usepackage{graphicx}
\graphicspath{{figs/}{supp_figs/}} 

\usepackage{physics}                
\usepackage{csquotes}               
\usepackage{multirow}               
\usepackage{booktabs}               

\usepackage[caption=false]{subfig}
\usepackage[colorlinks=true,citecolor=cyan,linkcolor=magenta,filecolor=magenta]{hyperref}
\usepackage[capitalize]{cleveref}

\usepackage{orcidlink}

\newcommand{\e}{\mathrm{e}}                 
\renewcommand{\i}{\mathrm{i}}               
\newcommand{\adjo}[1]{{#1}^{\dagger}}       
\newcommand{\nn}[1]{\langle #1 \rangle}     
\let\avec\vec
\renewcommand{\vec}[1]{{\vb{#1}}}           

\newcommand{\Ham}{\mathcal{H}}              
\newcommand{\BlochHam}{H}                   
\newcommand{\HopMat}{\mathsf{h}}            
\newcommand{\TG}{\Delta}                    
\newcommand{\tgel}{t}                       
\newcommand{\HTG}{\Gamma}                   
\newcommand{\htgel}{\gamma}                 
\newcommand{\HTGuc}{\HTGsc{1}}              
\newcommand{\HTGsc}[1]{\HTG^{(#1)}}         
\newcommand{\htg}[1]{\gamma_{#1}}           
\newcommand{\HPG}{G}                        
\newcommand{\HPGuc}{\HPG}                   
\newcommand{\gpres}[2]{\left.\left\langle #1\vphantom{#2} \right| #2 \right\rangle} 
\newcommand{\nsubg}{\trianglelefteq}        
\newcommand{\nsubgstr}{\vartriangleleft}    
\newcommand{\rtsgbusn}{\vartriangleright}   
\newcommand{\rtransv}[2]{T_{#1}\!(#2)}      
\newcommand{\bigunion}{\bigcup}             
\newcommand{\genus}{\mathfrak{g}}           
\newcommand{\ABZ}[1]{\mathrm{ABZ}^{(#1)}}   
\newcommand{\torus}[1]{\mathsf{T}^{#1}}     
\newcommand{\ccoord}{\mathsf{z}}

\newcommand{\gap}{\textsc{gap}}

\newcommand{\sectitle}[1]{\emph{{#1}.---}}

\newcommand{\tgquot}[2]{\mathrm{T}#1.#2}

\makeatletter
\def\maketitle{
\@author@finish
\title@column\titleblock@produce
\suppressfloats[t]}
\makeatother

\makeatletter
\def\bibsection{%
   \par
   \begingroup
    \baselineskip26\p@
    \bib@device{\hsize}{72\p@}%
   \endgroup
   \nobreak\@nobreaktrue
   \addvspace{19\p@}%
  }%
\makeatother

\renewcommand{\sectitle}[1]{\pdfbookmark[1]{#1}{#1}\emph{{#1}.---}}

\begin{document}

\title{Non-Abelian hyperbolic band theory from supercells}

\author{Patrick M. Lenggenhager\,\orcidlink{0000-0001-6746-1387}}
\email{plengg@pks.mpg.de}
\affiliation{Department of Physics, University of Zurich, Winterthurerstrasse 190, 8057 Zurich, Switzerland}
\affiliation{Condensed Matter Theory Group, Paul Scherrer Institute, 5232 Villigen PSI, Switzerland}
\affiliation{Institute for Theoretical Physics, ETH Zurich, 8093 Zurich, Switzerland}
\affiliation{Theoretical Physics Institute, University of Alberta, Edmonton, Alberta T6G 2E1, Canada}

\author{Joseph Maciejko\,\orcidlink{0000-0002-6946-1492}}
\email{maciejko@ualberta.ca}
\affiliation{Theoretical Physics Institute, University of Alberta, Edmonton, Alberta T6G 2E1, Canada}
\affiliation{Department of Physics, University of Alberta, Edmonton, Alberta T6G 2E1, Canada}

\author{Tom\'{a}\v{s} Bzdu\v{s}ek\,\orcidlink{0000-0001-6904-5264}}
\email{tomas.bzdusek@uzh.ch}
\affiliation{Department of Physics, University of Zurich, Winterthurerstrasse 190, 8057 Zurich, Switzerland}
\affiliation{Condensed Matter Theory Group, Paul Scherrer Institute, 5232 Villigen PSI, Switzerland}

\date{\today}

\begin{abstract}
Wave functions on periodic lattices are commonly described by Bloch band theory.
Besides Abelian Bloch states labeled by a momentum vector, hyperbolic lattices support non-Abelian Bloch states that have so far eluded analytical treatments.
By adapting the solid-state-physics notions of supercells and zone folding, we devise a method for the systematic construction of non-Abelian Bloch states.
The method applies Abelian band theory to sequences of supercells, recursively built as symmetric aggregates of smaller cells, and enables a rapidly convergent computation of bulk spectra and eigenstates for both gapless and gapped tight-binding models.
Our supercell method provides an efficient means of approximating the thermodynamic limit and marks a pivotal
step towards a complete band-theoretic characterization of hyperbolic lattices.
\end{abstract}

\maketitle

\sectitle{Introduction}
Hyperbolic lattices are uniform discretizations of the two-dimensional (2D) \emph{hyperbolic plane} with constant negative curvature.
Recent experimental realizations in metamaterials, including coplanar-waveguide resonator~\cite{Kollar:2019} and electric-circuit networks~\cite{Lenggenhager:2021}, have elevated them from objects of academic interest to building blocks for engineering metamaterials.
These advances have sparked a renewed interest in condensed-matter models on hyperbolic lattices, both in theory~\cite{Yu:2020,Boettcher:2020,Saa:2021,Zhu:2021,Stegmaier:2021,Ruzzene:2021,Zhu:2021,Bienias:2022,Urwyler:2022,Bzdusek:2022,Chen:2023,Mosseri:2022,Cheng:2022,Liu:2022,Tao:2022,Liu:2023,Basteiro:2022a,Basteiro:2022b,Gluscevich:2023} and experiment~\cite{Zhang:2022,Zhang:2023,Chen:2023a,Pei:2023,Chen:2023c}.
The fundamental construction involves \emph{regular tessellations}, where $q$ copies of regular $p\textrm{-}$gons meet at each vertex, denoted by $\{p, q\}$ in Schläfli notation, with $(p-2)(q-2)\,{>}\,4$.

For Euclidean lattices, Bloch's theorem labels Hamiltonian eigenstates by irreducible representations (IRs) of the translation group and enables a description in terms of a unit cell together with reciprocal space.
While Bloch's theorem has been generalized~\cite{Maciejko:2021,Maciejko:2022,Boettcher:2022,Ikeda:2021a,Ikeda:2021b,Kienzle:2022,Nagy:2022,Attar:2022} to hyperbolic lattices, this comes with fundamental difficulties.
First, Bloch's theorem requires periodic boundary conditions (PBC), but constructing finite PBC \emph{clusters} that approximate the thermodynamic limit is highly nontrivial~\cite{Maciejko:2022,Lux:2022,Lux:2023}.
Second, Euclidean translation groups are Abelian, such that only $1\textrm{D}\,\textrm{IRs}$ exist.
In contrast, hyperbolic translation groups admit higher-dimensional IRs; therefore, hyperbolic band theory (HBT) requires \emph{non-Abelian} Bloch states besides the usual Abelian ones~\cite{Maciejko:2022}.
We here refer to the approximation that considers only $1\textrm{D}\,\textrm{IRs}$ as \emph{Abelian} HBT (AHBT)~\cite{Maciejko:2021}.

To deal with these difficulties, various avenues have been explored.
Finite \emph{flakes} with open boundary conditions exhibit a macroscopic fraction of boundary sites, which is advantageous when interested in boundary effects~\cite{Urwyler:2022,Liu:2022,Tao:2022,Liu:2023}, but challenging when studying bulk properties.
Good agreement of AHBT with bulk-projected spectra on flakes is observed in some models~\cite{Urwyler:2022,Bzdusek:2022,Chen:2023}, but crucial features are missed in others~\cite{Tummuru:2023}.
Very recently, \textcite{Lux:2022,Lux:2023} have shown how to choose increasingly large PBC clusters to achieve convergence to the thermodynamic limit, while \textcite{Mosseri:2023} have computed the density of states (DOS) of gapless models using a continued-fraction method.
However, neither provides a reciprocal-space description, i.e., a description in terms of bulk states of the infinite lattice labeled by translation quantum numbers.

In this work, we introduce the \emph{supercell method} to gain systematic access to non-Abelian Bloch states using AHBT combined with particular sequences of PBC clusters.
We construct such sequences for various hyperbolic $\{p, q\}$ lattices using results from geometric group theory~\cite{Conder:2007}.
We observe rapid convergence of the DOS to the thermodynamic limit for various models.
Our approach is computationally more efficient than real-space methods and affords the conceptual advantages of labeling eigenstates by momenta.
We implement our algorithms in an open-source software package~\cite{HyperCells} for the computational algebra system \gap{}~\cite{GAP4}.

\sectitle{Supercells}
A lattice consists of copies of some chosen unit cell, generated by discrete translations forming a translation group $\HTG$.
While one commonly chooses a smallest \emph{primitive cell}, implying a maximal translation group $\HTGuc{}$, one can instead consider a \emph{supercell}, i.e., a collection of multiple primitive cells.
Accordingly, only a subgroup $\HTGsc{2}$ of translations $\HTGuc$ is required to generate the lattice.
An example pair of primitive cell and supercell of the (Euclidean) $\{4, 4\}$ lattice and the (hyperbolic) $\{8, 8\}$ lattice is illustrated in \cref{main:fig:supercell:lattice44,main:fig:supercell:lattice88}, respectively.

\begin{figure}[t]
    \subfloat{\label{main:fig:supercell:lattice44}}
    \subfloat{\label{main:fig:supercell:lattice88}}
    \subfloat{\label{main:fig:supercell:compact44}}
    \subfloat{\label{main:fig:supercell:compact88}}
    \subfloat{\label{main:fig:supercell:dos44}}
    \subfloat{\label{main:fig:supercell:dos88}}
    \centering
    \includegraphics[width=\linewidth]{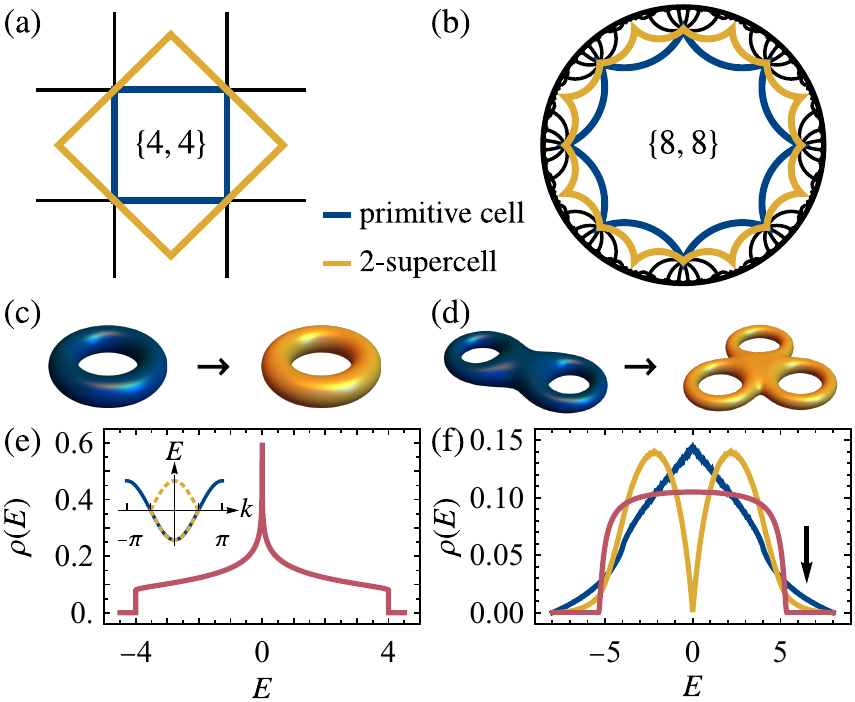}
    \caption{
        Supercell construction for (a,c,e) the Euclidean $\{4, 4\}$ and (b,d,f) hyperbolic $\{8, 8\}$ lattice.
        (a,b) Primitive cell (blue) and symmetrized \mbox{$2$-supercell} (yellow).
        (c,d) Compactified cells in real space.
        (e) Density of states $\rho$ of the nearest-neighbor hopping model on the $\{4, 4\}$ lattice as a function of energy $E$ showing the characteristic van Hove singularity. 
        The inset shows the momentum-space dispersion for the primitive cell (solid blue line) and for the supercell (yellow dashed line).
        (f) Density of Abelian Bloch states of the nearest-neighbor hopping model on the $\{8, 8\}$ lattice for the primitive cell (blue), for the $2$-supercell (yellow), and schematic extrapolation (for details see Supplemental Material~\cite{SM}) to large supercells (red). 
        The black arrow indicates a suppression near the band edges (see text).
    }
    \label{main:fig:supercell_construction}
\end{figure}

Dividing the lattice into copies of a chosen cell facilitates PBC, where the lattice is compactified on a closed manifold by identifying sides related by certain translations.
On Euclidean lattices, such \emph{PBC clusters} provide an approximation of the infinite lattice with well-converging bulk properties~\cite{Ashcroft:1976}.
To implement PBC on a single cell, opposite sides are identified and the cell is compactified on a torus---independent of its size (\cref{main:fig:supercell:compact44}).
In contrast, due to the negative curvature, hyperbolic PBC clusters are compactified on manifolds of genus $\genus{}\,{\geq}\,2$~\cite{Sausset:2007}.
According to the Riemann-Hurwitz formula~\cite{Miranda:1995}, the genus $\genus{}_\mathrm{sc}$ of a compactified supercell grows linearly with the number $N$ of primitive cells:
\begin{equation}
    \genus{}_\mathrm{sc} - 1 = N(\genus{}_\mathrm{pc}-1),
    \label{main:eq:RiemannHurwitz}
\end{equation}
where $\genus{}_\mathrm{pc}$ is the genus of the compactified primitive cell.
For the $\{8, 8\}$ lattice, the primitive cell is compactified on a \mbox{genus-$2$} surface, and the two-unit-cell supercell (\mbox{$2$-supercell}) on a \mbox{genus-$3$} surface (\cref{main:fig:supercell:compact88}).

Translation symmetry further enables a reciprocal-space description of the infinite lattice, considering not just a single PBC cluster but also all of its translation-related copies.
The choice of cell affects the reciprocal-space description.
To illustrate this, consider nearest-neighbor (NN) hopping models on the $\{4,4\}$ and $\{8,8\}$ lattices with Hamiltonian $\Ham=-\sum_{\nn{i,j}}\adjo{c_i}c_j$, where $\nn{i,j}$ denotes NNs.
For Euclidean lattices, the Brillouin zone (BZ) is reduced due to the enlargement of the cell, leading to band folding (\cref{main:fig:supercell:dos44}, inset);
nevertheless, the computed DOS is independent of the cell size.
By contrast, in the hyperbolic case, the density of Abelian Bloch states changes significantly when going from a primitive cell to a \mbox{$2$-supercell} (\cref{main:fig:supercell:dos88}).
However, below and in the Supplemental Material~\cite{SM}, we demonstrate that the DOS converges with increasing supercell size.

\begin{figure}
    \subfloat{\label{main:fig:triangle_group:lattice88}}
    \subfloat{\label{main:fig:triangle_group:generators}}
    \subfloat{\label{main:fig:triangle_group:lattice44}}
    \centering
    \includegraphics[width=\linewidth]{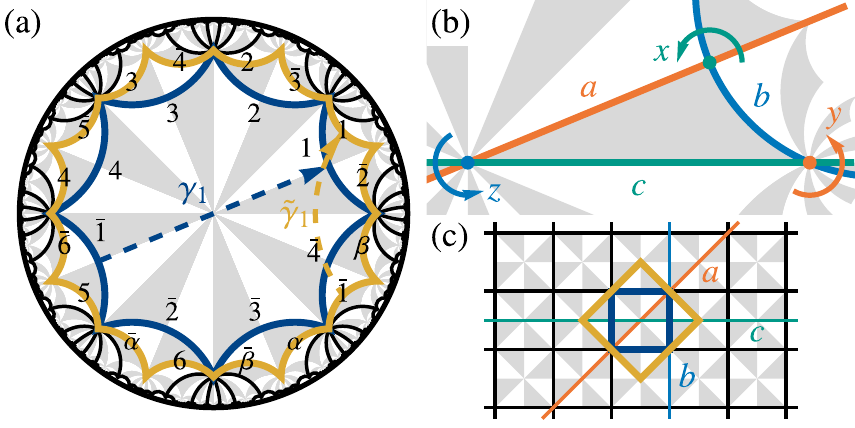}
    \caption{Symmetries of hyperbolic lattices. (a) $\{8,8\}$ lattice (black lines) with the triangle group $\TG(2,8,8)$ as space group (indicated by gray/white triangles). The primitive cell (blue polygon) and \mbox{$2$-supercell} (yellow polygon) and their edge identifications are shown: the edge $\bar{1}$ is related to $1$ by the translation generator $\htgel_1$ ($\tilde{\htgel}_1$ for the supercell). Edges related by composite translations are labeled by $\alpha$ and $\beta$~\cite{SM}. (b) Fundamental Schwarz triangle (gray) with reflections $a{,}\,b{,}\,c$ across the edges of the triangle and rotations $x\,{=}\,ab$, $y\,{=}\,bc$, $z\,{=}\,ca$ around the vertices. (c) Square lattice with primitive cell, supercell, triangle group $\TG(2,4,4)$, and reflection lines $a{,}\,b{,}\,c$.}
    \label{main:fig:triangle_group}
\end{figure}

\sectitle{Real-space perspective}
The symmetries of a $\{p,q\}$ lattice are captured~\cite{Magnus:1974,Boettcher:2022} by the \emph{triangle group} $\TG$ generated by reflections $a{,}\,b{,}\,c$ across the sides of a triangle, called \emph{Schwarz triangle}, with internal angles $\tfrac{2\pi}{r}$ (with $r\,{=}\,2$), $\tfrac{2\pi}{q}$, and $\tfrac{2\pi}{p}$ (\cref{main:fig:triangle_group:generators,main:fig:triangle_group:lattice44}). This is reflected in its \emph{presentation}
\begin{equation}
    \TG(r,q,p) = \gpres{a,b,c}{a^2,b^2,c^2,(ab)^r,(bc)^q,(ca)^p}
\end{equation}
with the \emph{relators}, appearing to the right of the vertical line, set to the identity.
Under the action of $\TG$, copies of the \emph{fundamental} Schwarz triangle $s_f$ tile the whole plane (see \cref{main:fig:triangle_group:lattice88,main:fig:triangle_group:lattice44} for the $\{8,8\}$ and $\{4,4\}$ lattices, respectively).
Formally, the abstract set $S$ of all Schwarz triangles is the orbit of $s_f$ under right action of $\TG:\,S\,{=}\,s_f\cdot \TG$.

Interpreted as a space group,  $\TG$ encompasses point-group operations and translations.
While the point group is generally not a subgroup of $\TG$ (even in Euclidean lattices), translations form a \emph{normal subgroup} $\HTG\,{\nsubgstr}\,\TG$~\cite{Bradley:1972}, i.e., any translation conjugated by a reflection or rotation is again a translation.
Indeed, $\HTG$ is usually defined~\cite{Boettcher:2022,Chen:2023} as the largest \emph{torsion-free} normal subgroup of orientation-preserving elements of $\TG$, where torsion-freeness captures the absence of elements of finite order in translation groups.
Since $\HTG\,{\nsubgstr}\,\TG$, the quotient $\TG/\HTG$ forms a group and plays the role of the point group.
The transversal $\rtransv{\TG}{\HTG}$ is a specific set of representatives of $\TG/\HTG$.

The choice of $\HTG$ defines the cell, which comprises a finite number of Schwarz triangles (\cref{main:fig:triangle_group:lattice88}) and therefore corresponds to a subset $C\,{\subset}\,S$ such that: (\emph{i})~none of the elements are related by translations, and (\emph{ii})~the right action of $\HTG$ on $C$ recovers $S$, i.e., $S\,{=}\,C\cdot\HTG$.
The coset decomposition $\TG=\bigunion_{\tgel\,{\in}\,\rtransv{\TG}{\HTG}} \HTG\tgel$ implies that $C\,{=}\,s_f\cdot\rtransv{\TG}{\HTG}$.
Different choices of $\rtransv{\TG}{\HTG}$ lead to cells $C$ differing in connectedness and symmetry.
Our algorithms~\cite{HyperCells} take $\Delta/\HTG{}$ as input, construct random and (for sufficiently small quotients) connected symmetric cells, and extract boundary identifications.

Our \emph{supercell method} is a natural and systematic way to form sequences of PBC clusters suited to a reciprocal-space interpretation.
We construct increasingly larger supercells, by recursively accreting smaller (super)cells in a symmetric fashion, starting with a single primitive cell.
This results in a nested sequence of finite-index normal subgroups,
\begin{equation}
    \HTGsc{1}\rtsgbusn \HTGsc{2} \rtsgbusn \dotsb \rtsgbusn \HTGsc{m}\rtsgbusn \dotsb,
    \label{main:eq:supercell-sequence}
\end{equation}
where $\HTGsc{m}\,{\nsubgstr}\,\Delta$ for all $m$ implies normality of the subgroup relationships in \cref{main:eq:supercell-sequence}. 
Although there is a unique plane-filling hyperbolic $\{p,q\}$ lattice, the PBC clusters can have different infinite-size limits~\cite{Maciejko:2022}, indicating the choice of sequence is crucial.
Recently, \textcite{Lux:2022,Lux:2023} proposed a similar condition with the additional constraint $\bigcap_{m\geq 1}^\infty\HTGsc{m}=\{1\}$ and argued that such sequences lead to a well-defined thermodynamic limit~\cite{Lueck:1994}.
Based on our results, we conjecture that the supercell sequences can be extended in a way that satisfies that additional constraint.
While the supercell method does not give a unique sequence, we anticipate that all valid sequences converge to the same limit, consistent with our observations~\cite{SM}.

Translation symmetry allows us to define hopping models by specifying only the hopping amplitudes $h^{uv}(\htgel)$ from site $v$ in the primitive cell $C^{(1)}$ to site $u$ in the primitive cell translated by $\htgel\,{\in}\,\HTGuc{}$.
[Here $C^{(m)}$ is the cell associated with the translation group $\HTGsc{m}$, and $N^{(m)}=|\Gamma^{(1)}/\Gamma^{(m)}|$ counts primitive cells in $C^{(m)}$.]
We additionally subdivide the lattice into copies of the \mbox{$N^{(m)}$-supercell}~$C^{(m)}$, so that the $N^{(m)}$ copies of $C^{(1)}$ in $C^{(m)}$ are generated by the quotient group $\HTGuc{}/\HTGsc{m}$.
By the coset decomposition, copies of the primitive cell are specified by $\eta_i\tilde{\htgel}$ with transversal elements $\eta_i\,{\in}\,\rtransv{\HTGuc{}}{\HTGsc{m}}$ and $\tilde{\htgel}\,{\in}\,\HTGsc{m}$, and the most general translation-invariant hopping model takes the form~\cite{SM},
\begin{equation}
    \Ham =\sum_{\substack{\tilde{\htgel},\tilde{\htgel}'\in\HTGsc{m}\\\eta_i,\eta_j\in\rtransv{\HTGuc{}}{\HTGsc{m}}}}
    \sum_{^{\phantom{\dagger}}u,v^{\phantom{\dagger}}}h^{uv}\left(\eta_{i}\tilde{\htgel}\tilde{\htgel}^{\prime-1}\eta_j^{-1}\right)
    {c_{\eta_i\tilde{\htgel}}^{u\;\;\dagger}}c_{\eta_j\tilde{\htgel}'}^v,
    \label{main:eq:Hamiltonian}
\end{equation}
where $\eta_{i}\tilde{\htgel}\tilde{\htgel}^{\prime-1}\eta_j^{-1}$ translates the primitive cell at $\eta_j\tilde{\htgel}'$ to that at $\eta_i\tilde{\htgel}$.
Our algorithms~\cite{HyperCells,HyperBloch} define hopping models on unit cells and extend models defined on a primitive cell to a supercell according to \cref{main:eq:Hamiltonian}.

\begin{figure}
    \centering
    \includegraphics[width=\linewidth]{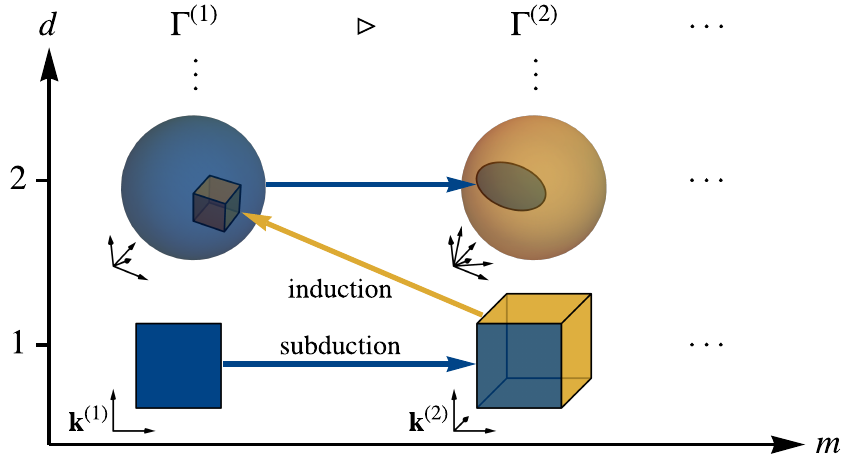}
    \caption{
        Illustration of the spaces of $d$-dimensional irreducible representations (IRs) of a sequence of translation subgroups $\HTGsc{m}$ corresponding to supercells with $N^{(m)}$ primitive cells.
        The spaces of $1\textrm{D}\,\textrm{IRs}$ are hypertori (illustrated as square and cube) with dimension growing linearly with $N^{(m)}$, while the spaces of higher-dimensional IRs are more complicated (illustrated as balls).
        The IRs of $\HTGsc{m}$ of dimension $d$ subduce (blue arrow) representations of $\HTGsc{m+1}$ of the same dimension and induce (yellow arrow) representations of $\HTGsc{m-1}$ of higher dimension (see text).
    }
    \label{main:fig:momentum_space}
\end{figure}

\sectitle{Reciprocal-space perspective}
In Euclidean lattices, translation symmetry constrains the form of Hamiltonian eigenstates via Bloch's theorem~\cite{Bloch:1929}.
Similarly, the \emph{automorphic Bloch theorem} for hyperbolic lattices~\cite{Maciejko:2022} stipulates that eigenstates $\psi_D$ of a translation-invariant Hamiltonian satisfy $\psi_D(\htgel^{-1}(\ccoord{}))=D(\htgel)\psi_D(\ccoord{})$, where $\htgel\,{\in}\,\HTG$ is a translation, $\ccoord{}$ the position coordinate, and $D$ an IR of $\HTG$.
By contrast with Euclidean lattices, $\HTG$ has IRs of dimensions $d\,{>}\,1$.
Nevertheless, we can block-diagonalize~\cite{SM} the Hamiltonian in \cref{main:eq:Hamiltonian} into blocks of Bloch Hamiltonians,
\begin{equation}
    \BlochHam(D) = \sum_{\tilde{\htgel}\in\HTGsc{m}}\HopMat(\tilde{\htgel})\otimes D(\tilde{\htgel}),
    \label{main:eq:Bloch-Hamiltonian}
\end{equation}
where $\HopMat_{ij}^{uv}(\tilde{\htgel})=h^{uv}(\eta_i\tilde{\htgel}\eta_j^{-1})$ is the hopping matrix within the supercell, and the unitary $(d\,{\times}\,d)$-matrix $D(\tilde{\htgel})$ generalizes the Bloch phase factor~\cite{Cheng:2022}.

Generally, no parametrization of the IRs $D$ is known, limiting a direct application of the automorphic Bloch theorem.
However, the space of $1\textrm{D}\,\textrm{IRs}$, the \emph{Abelian} BZ (ABZ), is well-understood: 
if the cell is compactified on a manifold of genus $\genus{}$, then $\mathrm{ABZ}$ is the $2\genus{}$-dimensional torus $\torus{2\genus{}}$ parametrized by momenta $\{0\leq k_i\,{<}\,2\pi\}_{i\,{=}\,1}^{2\genus{}}$ and the IRs are defined on the $2\genus{}$ generators $\htg{i}$ of $\HTG$ by $D_{\vec{k}}(\htg{i})=\e^{\i k_i}$~\cite{Maciejko:2021,Maciejko:2022,Boettcher:2022}.
While the ABZ sometimes is representative of the bulk spectrum of flakes~\cite{Urwyler:2022,Bzdusek:2022,Chen:2023,Chen:2023a}, important features can be missed~\cite{Mosseri:2023,Tummuru:2023}.
Studying the \emph{non-Abelian} Bloch states is therefore crucial for a complete reciprocal-space description.
Remarkably, as we explain below, AHBT applied to a sequence of supercells provides systematic access to non-Abelian Bloch states.

Considering the sequence of translation groups in \cref{main:eq:supercell-sequence}, each $\HTGsc{m}$ has a tower of $d$-dimensional IRs with $d\,{\geq}\,1$ (\cref{main:fig:momentum_space}).
However, due to the subgroup relationships, the IRs of different $\HTGsc{m}$ are \emph{not} independent.
First, the restriction of a $d$-dimensional IR of $\HTGsc{m}$ to its subgroup $\HTGsc{m+1}$ is a $d$-dimensional (possibly reducible) \emph{subduced} representation of $\HTGsc{m+1}$.
Second, any $d$-dimensional IR of $\HTGsc{m}$ implies a $(d\abs*{\HTGsc{m-1}/\HTGsc{m}})$-dimensional (possibly reducible) \emph{induced} representation of $\HTGsc{m-1}$~\cite{Fulton:1991}.
Thus, $1\textrm{D}\,\textrm{IRs}$ of $\HTGsc{m}$ subduce $1\textrm{D}\,\textrm{IRs}$ of $\HTGsc{m+1}$, but because $\ABZ{m+1}$ has larger dimension than $\ABZ{m}$, there must be $1\textrm{D}\,\textrm{IRs}$ of $\HTGsc{m+1}$ that induce higher-dimensional IRs of $\HTGsc{m}$.
Therefore, by studying the well-understood $1\textrm{D}\,\textrm{IRs}$ of supercells in the sequence, we gain access to a successively larger portion of higher-dimensional IRs of $\HTGuc{}$.
While this scheme does not reproduce \emph{all} IRs, we conjecture that it converges to the thermodynamic limit~\cite{Lux:2023} for $m\to\infty$.

\begin{figure}[t]
    \subfloat{\label{main:fig:dos_data:8kagome}}
    \subfloat{\label{main:fig:dos_data:83Haldane}}
    \centering
    \includegraphics[width=\linewidth]{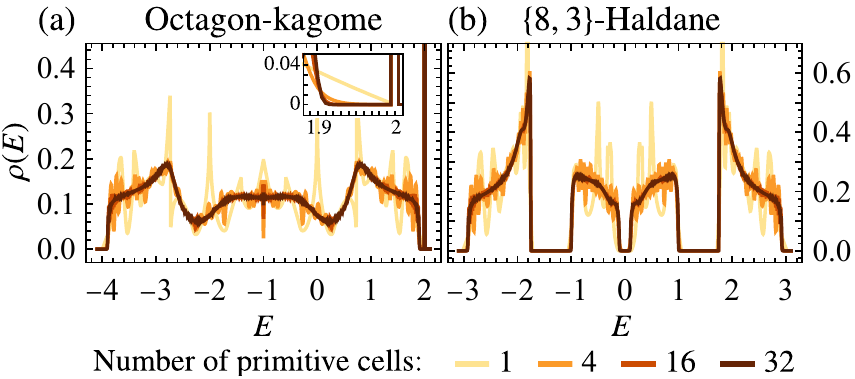}
    \caption{
        Density of states $\rho$ as a function of energy $E$ of (a) the nearest-neighbor (NN) model on the octagon-kagome lattice and (b) the Haldane model on the $\{8,3\}$ lattice with NN hopping $h_1\,{=}\,1$, next-NN hopping $h_2\,{=}\,1/6$, flux $\phi\,{=}\,\pi/2$, and sublattice mass $h_0\,{=}\,1/3$.
        The inset in (a) shows the depletion of $\rho(E)$ near the flat band at $E\,{=}\,2$.
    }
    \label{main:fig:dos_data}
\end{figure}

\sectitle{Results}
We illustrate the supercell method by computing the DOS for selected hopping models on hyperbolic lattices: the previously studied octagon-kagome~\cite{Bzdusek:2022} and $\{8,3\}\textrm{-}$Haldane~\cite{Urwyler:2022,Chen:2023} models, and a generalization to the $\{6,4\}$ lattice of the Benalcazar-Bernevig-Hughes (BBH) model~\cite{Benalcazar:2017}.
Each model is defined on a symmetric primitive cell and the DOS is computed~\cite{SM} by randomly sampling $\ABZ{m}$ in a sequence satisfying \cref{main:eq:supercell-sequence}.
We observe rapid convergence with system size: \cref{main:fig:dos_data:8kagome,main:fig:dos_data:83Haldane,main:fig:bbh:dos} show data for systems with only up to $32\,(768),\,32\,(512)$, and $64\,(1536)$ primitive cells (sites), respectively.
In sharp contrast, the DOS obtained from the corresponding PBC cluster (without applying AHBT) is extremely far from converged~\cite{SM}, demonstrating the computational power of our approach.

The NN hopping model on the octagon-kagome lattice has been analyzed in the context of flat bands in Ref.~\onlinecite{Bzdusek:2022}.
Using real-space arguments, the authors describe a band touching between the flat band and the dispersive bands.
In \cref{main:fig:dos_data:8kagome}, we observe that the DOS near the flat band is suppressed with increasing number of primitive cells $N$, suggesting that the gaplessness is a finite-size effect.
This DOS suppression is consistent with the expected behavior near a band edge.
Assuming a generic quadratic scaling of the energy dispersion with (Abelian) momentum near the band touching, $E\,{\propto}\,\vec{k}^2$, we obtain the DOS by integrating over the $2\genus{}$-dimensional ABZ: $\rho(E)\sim\int\dd[2\genus{}]\vec{k}\,\delta(E-v\vec{k}^2)\propto E^{\genus{}-1}$.
Since $\genus{}$ grows linearly with $N$ [\cref{main:eq:RiemannHurwitz}], this explains the DOS suppression near band edges, indicated in \cref{main:fig:supercell:dos88} and observed in all models (\cref{main:fig:dos_data,main:fig:bbh:dos}).

Next, we turn to the Haldane model on the $\{8,3\}$ lattice~\cite{Urwyler:2022,Chen:2023} which generalizes the original Haldane model on the honeycomb lattice~\cite{Haldane:1988}.
We adopt the parameter choices of Ref.~\onlinecite{Urwyler:2022} and show the converging DOS in \cref{main:fig:dos_data:83Haldane}.
Crucially, the characteristic DOS suppression near the edges of all three gaps indicates that the gaps obtained from AHBT are stable to the inclusion of non-Abelian Bloch states and are not caused by finite-size effects.

\begin{figure}[t]
    \subfloat{\label{main:fig:bbh:model}}
    \subfloat{\label{main:fig:bbh:dos}}
    \centering
    \includegraphics[width=\linewidth]{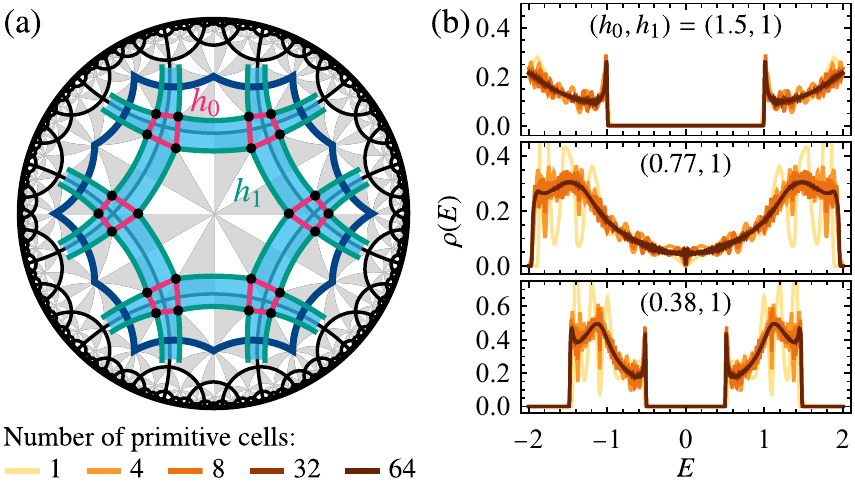}
    \caption{
        Benalcazar-Bernevig-Hughes model on the $\{6,4\}$ lattice.
        (a) Model definition on the primitive cell (blue polygon) of the $\{6,4\}$ lattice (black lines). There are four orbitals (black dots) at each site, coupled by inter-site hoppings $h_0$ (magenta) and intra-site hoppings $h_1$ (green). Light blue shading of plaquettes bounded by magenta/green lines indicates $\pi$-fluxes.
        (b) Density of states for the indicated choices of $(h_0,h_1)$. The top/middle/bottom subpanel corresponds to the trivial/critical/nontrivial phase.
    }
    \label{main:fig:bbh}
\end{figure}

Finally, motivated by the recent interest in higher-order topological phenomena on hyperbolic flakes~\cite{Tao:2022,Liu:2023}, we introduce the BBH model on the $\{6,4\}$ lattice.
Similar to its original version on the square lattice~\cite{Benalcazar:2017}, the model is defined on a lattice with fourfold coordination, has four orbitals per site, and exhibits $\pi$-fluxes through the quadrilateral plaquettes (\cref{main:fig:bbh:model}).
The intra-site hopping $h_0$ may differ from the inter-site hopping $h_1$.
As in the Euclidean case, this arrangement leads to a \emph{trivial} phase for $\abs{h_0}\,{\gg}\,\abs{h_1}$ with effectively independent rings centered at lattice sites, and a \emph{nontrivial} phase for $\abs{h_0}\ll\abs{h_1}$ with effectively independent rings centered on the plaquettes.
The computed DOS for the two phases and the transition are shown in \cref{main:fig:bbh:dos}.
The trivial and the nontrivial phase both exhibit an energy gap that remains stable when going to larger supercells.
The gap closes at $h_0/h_1\,{\approx}\,0.77$, indicating a phase transition.
Interestingly, for small supercells the transition appears semimetallic with vanishing DOS at $E\,{=}\,0$.
However, this is a finize-size effect and the DOS ultimately converges to a finite value, implying a \emph{metallic} transition.

\sectitle{Conclusions}
We have introduced a method for systematically studying non-Abelian Bloch states in hyperbolic lattices by applying Abelian hyperbolic band theory to sequences of supercells, in analogy to zone folding in solid-state physics.
This provides a substantial step toward a complete reciprocal-space description, which we believe to be consistent with recent work~\cite{Lux:2022,Lux:2023} in real space.
While real-space methods scale suboptimally due to the increasing number of noncontractible loops~\cite{Mosseri:2023}, the combination of real-space supercells with reciprocal-space momenta in our approach appears to mitigate this problem and additionally provides true bulk states instead of finite-size approximations.
Our DOS results on gapless elementary nearest-neighbor models are in agreement with previous results obtained using a different method~\cite{Mosseri:2023}, but we additionally studied topological models exhibiting energy gaps.
Our method has three substantial advantages over Ref.~\onlinecite{Mosseri:2023}: (\emph{i}) it gives direct access to bulk eigenstates, (\emph{ii}) it allows for parallel computation through separating the Hilbert space into $\vec{k}$-sectors, and (\emph{iii}) there is no extra computational cost for longer-range hoppings.

Looking ahead, we anticipate our reciprocal-space supercell method will facilitate advances in HBT such as symmetry analysis~\cite{Chen:2023}, low-energy expansions~\cite{Tummuru:2023}, and topological band theory, including the recently studied 2D hyperbolic model~\cite{Zhang:2023} with nontrivial second Chern number~\cite{Tummuru:2023}.
Developing systematic algorithms for generating longer sequences of $\Delta/\Gamma^{(m)}$ quotients beyond those in Ref.~\onlinecite{Conder:2007} would be beneficial for achieving better convergence.
Finally, we hope the implementation of our approach in a publicly available software package~\cite{HyperCells,HyperBloch,SDC} will accelerate further studies of hyperbolic quantum matter.

\let\oldaddcontentsline\addcontentsline     
\renewcommand{\addcontentsline}[3]{}        

\sectitle{Acknowledgments}
We would like to thank I.~Boettcher, A.~Chen, A.~Stegmaier, L.~K.~Upreti, S.~Dey, T.~Neupert, E.~Prodan, and T.~Tummuru for valuable discussions and R.~Mosseri and J.~Vidal for sharing their data, allowing a quantitative comparison of our results.
P.~M.~L.~and T.~B.~were supported by the Ambizione grant No.~185806 by the Swiss National Science Foundation (SNSF). 
P.~M.~L.~is grateful for the hospitality of the Theoretical Physics Institute at the University of Alberta, where part of this work was completed.
T.~B.~was supported by the Starting Grant No.~211310 by SNSF.
J.~M.~was supported by NSERC Discovery Grants \#RGPIN-2020-06999 and \#RGPAS-2020-00064; the Canada Research Chair (CRC) Program; the Government of Alberta's Major Innovation Fund (MIF); the Tri-Agency New Frontiers in Research Fund (NFRF, Exploration Stream); and the Pacific Institute for the Mathematical Sciences (PIMS) Collaborative Research Group program.

\pdfbookmark[1]{References}{references}
\bibliography{main}
\onecolumngrid

\let\addcontentsline\oldaddcontentsline     

\clearpage

\setcounter{page}{1}
\setcounter{equation}{0}
\setcounter{section}{0}
\setcounter{figure}{0}

\renewcommand{\theequation}{S\arabic{equation}}
\renewcommand{\thefigure}{S\arabic{figure}}
\renewcommand{\theHfigure}{S\arabic{figure}}
\renewcommand{\thetable}{S\arabic{table}}

\makeatletter 
    
\renewcommand\onecolumngrid{%
\do@columngrid{one}{\@ne}%
\def\set@footnotewidth{\onecolumngrid}%
\def\footnoterule{\kern-6pt\hrule width 1.5in\kern6pt}%
}

\renewcommand\twocolumngrid{%
        \def\footnoterule{
        \dimen@\skip\footins\divide\dimen@\thr@@
        \kern-\dimen@\hrule width.5in\kern\dimen@}
        \do@columngrid{mlt}{\tw@}
}%

\makeatother  

\title{Supplemental Material for:\texorpdfstring{\\}{ }Non-Abelian hyperbolic band theory from supercells}

\maketitle
\onecolumngrid

\tableofcontents

\section{Hyperbolic tight-binding models}\label{Sec:HTB}

Generically, a tight-binding model is defined in real space by orbitals at certain positions with nontrivial on-site symmetry group, so-called \emph{Wyckoff positions}, and couplings between them.
Thus, the first step in describing hyperbolic tight-binding models is to find and enumerate these Wyckoff positions.
In \cref{Sec:HTB:Wyckoff} we describe how to label the Wyckoff positions based on the triangle group, first in the infinite lattice, then on the lattice subdivided into unit cells, and finally on the lattice subdivided into supercells, which themselves are subdivided into primitive cells.
Next, in \cref{Sec:HTB:Ham}, we discuss generic forms of hopping Hamiltonians defined on infinite lattices and on finite clusters with periodic boundary conditions (PBC).
Finally, in \cref{Sec:HTB:Bloch-Ham}, we show how to block-diagonalize the Hamiltonian in terms of blocks of generalized Bloch Hamiltonians.

\subsection{Algebraic labeling of Wyckoff positions and unit cells}\label{Sec:HTB:Wyckoff}

To define tight-binding models in real space, we first need to develop a systematic method for labeling the potential positions of orbitals, i.e., the lattice sites.
These sites are typically located at high-symmetry positions, i.e., points with nontrivial on-site symmetry group, and depend on the underlying space group (the triangle group).
Here, we limit our attention to the maximally symmetric positions, i.e., those whose site-symmetry groups are not a proper subgroup of any other site-symmetry group~\cite{Cano:2020} (which for the regular $\{p,q\}$ tessellations correspond to centers of rotation symmetries), arriving at \cref{eq:all-sites}.

Next, we discuss how to subdivide the infinite lattice into unit cells and identify a set of representatives of all high-symmetry positions contained within the unit cell; these are the (maximally symmetric) Wyckoff positions, cf.~\cref{eq:unit-cell-Wyckoff-positions}.
That allows us to relabel all sites in the infinite lattice in terms of the unit cell they are contained in and the corresponding Wyckoff position, cf.~\cref{eq:Wyckoff-positions}.

Finally, we add two additional levels of subdivision: instead of covering the lattice directly with copies of a primitive cell, we subdivide it into large \emph{super-supercells} composed of supercells, which themselves are composed of primitive cells. 
The motivation for doing this is that we may wish to consider large but finite PBC clusters instead of the infinite hyperbolic lattice. 
In this language, a finite PBC cluster can be described as a single super-supercell.
In general, the primitive cells covering the infinite lattice are labeled according to \cref{eq:super-supercell-decomp} in terms of which super-supercell, which supercell within the super-supercell, and which primitive cell within the supercell they lie in.

\subsubsection{Tiling by Schwarz triangles}
Recall from the main text that the triangle group
\begin{equation}
    \TG(r,q,p) = \gpres{a,b,c}{a^2,b^2,c^2,(ab)^r,(bc)^q,(ca)^p}
\end{equation}
acts as a space group and tiles the plane by Schwarz triangles, cf.~\cref{main:fig:triangle_group}.
By defining the right group action of $\TG$ on the abstract set $S$ of all Schwarz triangles, $S$ can be viewed as generated by acting with all elements of $\TG$ on one particular element of $S$, the fundamental Schwarz triangle (FST) $s_f$: $S=s_f\cdot\TG$, cf.~\cref{main:fig:triangle_group:generators}.

The vertices of the Schwarz triangles in the tiling form a hyperbolic triangular lattice where each site has a nontrivial on-site symmetry group generated by the reflection symmetries with mirror lines passing through the vertex, cf.~\cref{main:fig:triangle_group:generators}.
Here, we find it more convenient to work directly with the proper triangle group
\begin{equation}
    \TG^+(r,q,p) = \gpres{x,y,z}{xyz,x^r,y^q,z^p},
\end{equation}
with the embedding homomorphism $\TG^+\to\TG$ defined by $x\mapsto ab$, $y\mapsto bc$, $z\mapsto ca$.
The proper subgroup of the above-mentioned on-site symmetry group of a given vertex is then generated by the rotation around that vertex, cf.~\cref{main:fig:triangle_group:generators}.
The set $V$ of all vertices can be divided into three subsets $V_z$, $V_y$, $V_x$ according to which vertex of the FST they originate from under action of $\TG$.
Thus, we have $V=V_z\cup V_y\cup V_x$.
Below we discuss $V_z$, but the discussion for the other two sets is analogous.

The on-site symmetry group of a vertex in $V_z$ is
\begin{equation}
    \TG_z^+ = \gpres{z}{z^p}
\end{equation}
and is exactly the stabilizer of that vertex if we define a right action of $\TG^+$ on the abstract set $V_z$.
By the orbit-stabilizer theorem, there exists a bijection between the orbit $v_z\cdot\TG^+$ of some $v_z\in V_z$ (assumed to be a vertex of the FST) and the quotient $\TG^+/\TG_z^+$, or equivalently between $V_z$ and the right transversal of $\TG_z^+$ in $\TG^+$
\begin{equation}
    V_z \cong \rtransv{\TG^+}{\TG_z^+}.
\end{equation}
Thus, the vertices are labelled by $(w,[t]_w)$ for $w\in\{x,y,z\}$ and $[t]_w\in\TG^+/\TG_w^+$ where a set of distinct cosets $[t]$ is given by a choice of representatives, i.e., the right transversal $\rtransv{\TG^+}{\TG_w^+}$ mentioned above:
\begin{equation}
    V \cong \left\{(w,[t]_w)\,:\,w\in\{x,y,z\}, t\in \rtransv{\TG^+}{\TG_w^+}\right\}.
    \label{eq:all-sites}
\end{equation}
Note that for the orbit-stabilizer theorem it is important to pair right action with right cosets (and right transversal).
Alternatively, we can work with the full groups $\TG$ and $\TG_z$ to arrive at the same conclusion: $V_z\cong\rtransv{\TG}{\TG_z}$, because $\TG$ is the semidirect product of $\TG^+$ with $\mathbb{Z}_2$, the group generated by one of the reflections.

\subsubsection{Bravais lattice and unit cells}
The lattice formed by the vertices of the Schwarz triangles does not generally form a Bravais lattice.
To obtain the Bravais lattice corresponding to a triangle group $\TG(r,q,p)$, a translation group $\HTG\nsubgstr\TG^+$ needs to be identified as discussed in the main text.
The translation group then defines the subdivision of the infinite lattice into copies of the unit cell via the quotient group $\TG/\HTG$ (which is a group because $\HTG\nsubgstr\TG$).
Formally, we consider the decomposition of $\TG$ into (disjoint) right cosets of $\HTG$ (which are equal to the left cosets, because $\HTG\nsubgstr\TG$) using the transversal $T_\Delta(\Gamma)$:
\begin{equation}
    \TG = \bigcup_{g_j\in\rtransv{\TG}{\HTG}} \HTG g_j = \bigcup_{g_j\in\rtransv{\TG}{\HTG}} g_j\HTG = \bigcup_{\htgel\in\HTG}\rtransv{\TG}{\HTG}\htgel,
\end{equation}
such that the set $S$ of all Schwarz triangles is decomposed as
\begin{equation}
    S = s_f\cdot\TG = \bigcup_{\htgel\in\HTG}s_f\cdot\left(\rtransv{\TG}{\HTG}\htgel\right) = \bigcup_{\htgel\in\HTG}\left(s_f\cdot\rtransv{\TG}{\HTG}\right)\cdot\htgel,
\end{equation}
i.e., $S$ is divided into copies of $C=s_f\cdot\rtransv{\TG}{\HTG}$ under action of the translation group:
\begin{equation}
    S = \bigcup_{\htgel\in\HTG}C\cdot\htgel = C\cdot\HTG,
\end{equation}
where $C$ is interpreted as the set of Schwarz triangles in the unit cell.
Above, we have emphasized the fact that $\rtransv{\TG}{\HTG}$ is a set of chosen representatives of the cosets in $\TG/\HTG$ by labeling elements by $g_j$ with the index $j$ enumerating the representatives of different cosets.
The choice of $\rtransv{\TG}{\HTG}$ determines the unit cell in real space (relative to the FST), e.g., whether it is a connected region and whether it is symmetric under point-group operations.
The software package~\cite{HyperCells} accompanying our work includes algorithms that choose $\rtransv{\TG}{\HTG}$ such that $C$ is connected and symmetric, given as input $\TG^+/\HTG$ and $w\in\{x,y,z\}$ specifying the center and therefore symmetry of the unit cell.

By the isomorphism theorem, the quotient $\TG/\HTG$ is isomorphic to the group $\HPG$, which we interpret as the point group, since it maps the unit cell $C$ to itself.
Using this isomorphism, we can define the induced right action from $G$ on $C$, such that $C=s_f\cdot G$.
To label vertices, working with the proper subgroups is again more convenient (but not necessary).
Note that the semidirect product structure of $\TG$ implies $\HPG^+\cong\TG^+/\HTG$.
Further, for any $w\in\{x,y,z\}$, $\TG_w^+$, by definition in terms of a single rotation generator, does not have any torsion-free elements, such that the stabilizer $\HPG_w^+$ of $v_w$ (modulo translations in $\HTG$) under the right action of $G^+$ is isomorphic to the full stabilizer: $\HPG_w^+\cong\TG_w^+$.
This now allows us to label the vertices in the unit cell, i.e., the Wyckoff positions: $V^C = V_z^C\cup V_y^C\cup V_x^C$ with $V_w^C \cong \rtransv{\HPG^+}{\HPG_w^+}$, i.e.,
\begin{equation}
    V^C \cong \left\{(w,[\delta]_w)\,:\,w\in\{x,y,z\}, \delta\in \rtransv{\HPG^+}{\HPG_w^+}\right\},
    \label{eq:unit-cell-Wyckoff-positions}
\end{equation}
where $[\delta]_w\in\HPG^+/\HPG_w^+$.
Note that the subscript again indicates the quotient group in which the coset lies.
We now want to write \emph{all} vertices, i.e., the infinite set $V$, in terms of translated Wyckoff positions, i.e., in terms of the finite set $V^C$ and translations in $\HTG$.

By the coset decomposition
\begin{equation}
    \TG^+ = \bigcup_{g_j\in\rtransv{\TG^+}{\HTG}} g_j\HTG,
\end{equation}
any $\tgel\in\TG^+$ can be written as
\begin{equation}
    \tgel = g_j\htgel',
\end{equation}
for unique $g_j\in\rtransv{\TG^+}{\HTG}$ and $\htgel'\in\HTG$.
Using the quotient group isomorphism $i\,:\,\TG^+/\HTG\to \HPG^+$, the image $i([g_j])$ of the unique coset $[g_j]\in\TG^+/\HTG$ associated with $g_j\in\rtransv{\TG^+}{\HTG}$ can be rewritten using the right coset decomposition (for any choice of $w\in\{x,y,z\}$):
\begin{equation}
    \HPG^+ = \bigcup_{\delta_k\in\rtransv{\HPG^+}{\HPG_w^+}} \HPG_w^+\delta_k,
\end{equation}
i.e.,
\begin{equation}
    i([g_j]) = i([w^n])\delta_k
\end{equation}
with unique $n\in \{0,1,\dotsc,|\HPG_w^+|-1\}$ and $\delta_k\in\rtransv{\HPG^+}{\HPG_w^+}$.
Note that $n=n(j)$ and $k=k(j)$, i.e., they implicitly depend on the specific choice of $j$, but we suppress this dependence for better readability.
Applying the inverse of $i$ and recognizing that $i^{-1}(\delta_k)=[g]$ for a unique $g\in\rtransv{\TG^+}{\HTG}$, which we denote by $g_{u_w}$ to emphasize that $i([g_{u_w}])\in\rtransv{\HPG^+}{\HPG_w^+}$, results in
\begin{equation}
    [g_j] = [w^n][g_{u_w}],
\end{equation}
which in turn implies (due to $\HTG\nsubg\TG^+$)
\begin{equation}
    g_j = w^ng_{u_w}\htgel'',
\end{equation}
for a unique $\htgel''\in\HTG$.
Thus, we have shown that for any choice $w\in\{x,y,z\}$, an arbitrary $\tgel\in\TG^+$ can be written as
\begin{equation}
    \tgel = w^ng_{u_w}\htgel
    \label{eq:Wyckoff-decomp}
\end{equation}
for a unique $n\in \{0,1,\dotsc,|\HPG_w^+|-1\}$, $g_{u_w}\in\rtransv{\TG^+}{\HTG}$ with $i([g_{u_w}])\in\rtransv{\HPG^+}{\HPG_w^+}$, and $\htgel\in\HTG$.
Note that the order of the elements in \cref{eq:Wyckoff-decomp} is important (and is influenced by the fact that we use right action).
The rotation $w^n$ is to the very left, as required by the orbit-stabilizer theorem; when generating the set $V$ by right action from the three vertices $v_w$, $w\in\{x,y,z\}$, $w^n$ leaves $v_w$ invariant due to being in its stabilizer $\TG_w^+$.
The order of $g_{u_w}$ and $\htgel$, on the other hand, is physically motivated (mathematically, the opposite order would be equally valid, since $\HTG\nsubgstr\TG^+$ implies that $g_{u_w}\htgel$ could be written with \emph{different} translation to the left of $g_{u_w}$ instead): $g_{u_w}$ specifies the position of the vertex in each unit cell and then the \emph{whole} unit cell containing all vertices is translated by $\htgel$.

This allows us to label all Wyckoff positions in the unit cell by $u=(w,[g_{u_w}]_w)$ where $[g_{u_w}]_w\in\rtransv{\HPG^+}{\HPG_w^+}$.
Correspondingly, Wyckoff positions in the infinite lattice can be labelled by $(u,\htgel)$, where $\htgel\in\HTG$ labels the unit cell, i.e.,
\begin{equation}
    V \cong \bigcup_{\htgel\in\HTG}\left\{(u,\htgel)\,:\,u = (w,[g_{u_w}]_w), w\in\{x,y,z\},g_{u_w}\in\rtransv{\TG^+}{\HTG},i([g_{u_w}])\in\rtransv{\HPG^+}{\HPG_w^+}\right\}.
    \label{eq:Wyckoff-positions}
\end{equation}
More practically, determining the set $V$ of distinct vertices requires to first choose $\rtransv{\TG^+}{\HTG}\subset\TG^+$ and then find the subset of elements $g_{u_w}$ satisfying $i([g_{u_w}])\in\rtransv{\HPG^+}{\HPG_w^+}$.

The software package~\cite{HyperCells} accompanying our work includes algorithms that, given $\HPG^+$ and $w$ as input, construct (1) a list of Wyckoff positions in the unit cell labelled by $u=(w,u_w)$, (2) the compactified nearest-neighbor graph formed by them, including corresponding translations $\htgel$ for edges crossing the unit cell boundary, and (3) a minimal list of translation generators chosen from those to ensure that they are translations to adjacent unit cells.

Part of the output of those algorithms for two quotient groups of $\TG(2,8,8)$ is illustrated in \cref{main:fig:triangle_group:lattice88}.
The set $S$ of Schwarz triangles is shown as gray/white triangles, the subset $C^{(1)}$ corresponding to the primitive cell is indicated by the blue polygon and the subset $C^{(2)}$ corresponding to the first supercell by the orange polygon.
The compactified nearest-neighbor graphs are not shown explicitly but can be inferred: edges go along sides of the Schwarz triangles and for a given cell (primitive or supercell) edges on the cell boundary are identified according to the translations that relate them.
These translations can be either generators $\htg{i}$ of the corresponding translation group (indicated by the corresponding index $i$), their inverses $\htg{i}^{-1}$ (indicated by $\bar{i}$), or composite translations and their inverses.
Specifically for the $2$-supercell in \cref{main:fig:triangle_group:lattice88} there are two pairs of boundary edges related by the composite translations: $\alpha=\tilde{\htg{5}}^{-1}\tilde{\htg{3}}^{-1}\tilde{\htg{1}}^{-1}$ and $\beta=\tilde{\htg{6}}^{-1}\tilde{\htg{4}}^{-1}\tilde{\htg{2}}^{-1}$.

\subsubsection{Supercells}
Above, we have introduced labels for Wyckoff positions and unit cells.
Next, we consider supercells made up of primitive cells.
We assume a normal sequence of translation groups $\HTGsc{m}\nsubgstr\TG^+$, $m\geq 1$, satisfying $\HTGsc{m+1}\nsubgstr\HTGsc{m}$, as discussed in the main text.
The primitive cells $C^{(1)}$ within the infinite lattice are labeled by $\htgel^{(1)}\in\HTGuc{}$.
Because $\HTGsc{m}$ is a normal subgroup of $\HTGuc{}$, we can once more consider the coset decomposition
\begin{equation}
    \HTGsc{1} = \bigcup_{\eta_i^{(1)}\in\rtransv{\HTGuc{}}{\HTGsc{m}}}\eta_i^{(1)}\HTGsc{m},
\end{equation}
which implies that any $\htgel^{(1)}\in\HTGuc{}$ can be decomposed into unique $\eta_i^{(1)}\in\rtransv{\HTGuc{}}{\HTGsc{m}}$ and $\htgel^{(m)}\in\HTGsc{m}$:
\begin{equation}
    \htgel^{(1)} = \eta_i^{(1)}\htgel^{(m)},
    \label{eq:supercell-decomp}
\end{equation}
where $\eta_i^{(1)}\in\rtransv{\HTGuc{}}{\HTGsc{m}}$ now labels the primitive cells $C^{(1)}$ within a supercell $C^{(m)}$ and $\htgel^{(m)}\in\HTGsc{m}$ the supercells $C^{(m)}$ in the infinite lattice.
Let us again comment on the order of group elements.
The same argument as for the position of $\htgel$ in \cref{eq:Wyckoff-decomp} applies here: $\eta_i^{(1)}$, which describes the internal structure of the supercell, appears to the left of the translation $\htgel^{(m)}$ connecting different copies of the supercell.
With that order, if two primitive cells in some supercell are related by a translation generator $\eta_i^{(1)}$, the corresponding copies of the primitive cells in the translated supercell are still related by $\eta_i^{(1)}$.

We now consider a larger super-supercell $C^{(n)}$ given by $\HTGsc{n}$ with $n>m$.
Such a super-supercell consists of copies of the smaller supercell $C^{(m)}$, which in turn consists of copies of the primitive cell $C^{(1)}$.
Using the fact that for $\htgel^{(m)}\in\HTGsc{m}$ there exist unique $\eta_{i'}^{(m)}\in\rtransv{\HTGsc{m}}{\HTGsc{n}}$ and $\htgel^{(n)}\in\HTGsc{n}$ such that $\htgel^{(m)}=\eta_{i'}^{(m)}\htgel^{(n)}$, we find that any $\htgel^{(1)}\in\HTGuc{}$ can be decomposed into unique $\eta_i^{(1)}\in\rtransv{\HTGuc{}}{\HTGsc{m}}$, $\eta_{i'}^{(m)}\in\rtransv{\HTGsc{m}}{\HTGsc{n}}$ and $\htgel^{(n)}\in\HTGsc{n}$:
\begin{equation}
    \htgel^{(1)} = \eta_i^{(1)}\eta_{i'}^{(m)}\htgel^{(n)}.
    \label{eq:super-supercell-decomp}
\end{equation}
We consistently label elements of $\HTGsc{m}$ with the corresponding superscript, e.g., $\htgel^{(m)}\in\HTGsc{m}$ and $\eta_{i'}^{(m)}\in\HTGsc{m}$, where the subscript indicates that $\eta_{i'}^{(m)}$ is an element of a transversal, in this case of $\rtransv{\HTGsc{m}}{\HTGsc{n}}$.
Note that $\HTGsc{m}\subset\HTGuc{}$, such that all elements in \cref{eq:super-supercell-decomp} are elements of $\HTGuc{}$.

\subsection{Hopping Hamiltonian}\label{Sec:HTB:Ham}

In this subsection, we develop the formalism for defining hopping Hamiltonians on both infinite hyperbolic lattices as well as PBC clusters, based on the algebraic description of the lattice introduced in the previous subsection.
We start by subdividing the infinite lattice into primitive cells, each of which contains orbitals at certain Wyckoff positions.
This utilizes the decomposition given in \cref{eq:Wyckoff-decomp,eq:Wyckoff-positions} and results in \cref{eq:hopp-Ham:primitive-cells}.
Next, we further subdivide the lattice into supercells $C^{(m)}$ consisting of several primitive cells.
Based on the decomposition in \cref{eq:supercell-decomp}, this results in \cref{eq:hopp-Ham:m-supercell}.
Finally, we add one more layer of subdivision: larger super-supercells $C^{(n)}$ consisting themselves of several copies of $C^{(m)}$.
This last step reflects the decomposition in \cref{eq:super-supercell-decomp} and results in \cref{eq:hopp-Ham:n-supercell}.
Instead of working on the infinite lattice, one can consider finite PBC clusters: the PBC identify each site outside of the cluster with a particular site inside it.
This allows us to immediately write down the Hamiltonians on PBC clusters formed by \emph{a single} supercell $C^{(m)}$ or a single super-supercell $C^{(n)}$, in \cref{eq:Ham-m-PBC,eq:Ham-n-PBC}, respectively.

A generic hopping Hamiltonian defined on the primitive cell $C^{(1)}$ with translation group $\HTGuc{}$ and hopping amplitude $h^{uv}(\htgel)$ for hopping from orbital $v$ in the unit cell to orbital $u$ in the unit cell translated by $\htgel$ is given by
\begin{equation}
    \Ham = \sum_{\htgel^{(1)},\breve{\htgel}^{(1)}\in\HTGuc{}}\sum_{u,v} h^{uv}\left(\htgel^{(1)}\mbox{$\breve{\htgel}^{(1)}$}^{-1}\right){c^u_{\htgel^{(1)}}}^{\hspace{-1mm}\dagger\hspace{1mm}}c^v_{\breve{\htgel}^{(1)}},
    \label{eq:hopp-Ham:primitive-cells}
\end{equation}
where $\adjo{c^u_{\htgel}}$ is the creation operator for the orbital $u$ in the unit cell $\htgel$.
Note that the more generic amplitude $h^{uv}(\htgel^{(1)},\breve{\htgel}^{(1)})$ simplifies to $h^{uv}(\htgel^{(1)}\mbox{$\breve{\htgel}^{(1)}$}^{-1})$ due to translation invariance.

If we further subdivide the infinite lattice into supercells $C^{(m)}$ with translation group $\HTGsc{m}$, then the same Hamiltonian can be rewritten as
\begin{equation}
    \Ham = \sum_{\htgel^{(m)},\breve{\htgel}^{(m)}\in\HTGsc{m}}\sum_{\eta_i^{(1)},\eta_j^{(1)}\in\rtransv{\HTGuc{}}{\HTGsc{m}}}\sum_{u,v} h^{uv}\left(\eta_i^{(1)}\htgel^{(m)}\mbox{$\breve{\htgel}^{(m)}$}^{-1}\mbox{$\eta_j^{(1)}$}^{-1}\right){c^u_{\eta_i^{(1)}\htgel^{(m)}}}^{\hspace{-2mm}\dagger\hspace{2mm}}c^v_{\eta_j^{(1)}\breve{\htgel}^{(m)}},
    \label{eq:hopp-Ham:m-supercell}
\end{equation}
where we used the unique decomposition given in \cref{eq:supercell-decomp}.
With simplified notation $\htgel^{(m)}\mapsto\tilde{\htgel}$, $\breve{\htgel}^{(m)}\mapsto\tilde{\htgel}'$, and $\eta_i^{(1)}\mapsto\eta_i$, this is exactly the Hamiltonian appearing in \cref{main:eq:Hamiltonian}.
Restricting to a single copy of the supercell implies that sites related by some $\htgel^{(m)}\in\HTGsc{m}$ are identified.
Formally, this is implemented by specifying the primitive cell not by elements of $\HTGuc{}$, but by cosets $[\eta_i^{(1)}]_{(1,m)}\in\HTGuc{}/\HTGsc{m}$, where $\eta_i^{(1)}\in\rtransv{\HTGuc{}}{\HTGsc{m}}$.
Here, we have introduced the subscript $(1,m)$ to indicate the type of coset.
This results in the Hamiltonian for a finite PBC cluster:
\begin{equation}
    \Ham_\text{$m$-PBC} = \sum_{\htgel^{(m)}\in\HTGsc{m}}\sum_{\eta_i^{(1)},\eta_j^{(1)}\in\rtransv{\HTGuc{}}{\HTGsc{m}}}\sum_{u,v} h^{uv}\left(\eta_i^{(1)}\htgel^{(m)}\mbox{$\eta_j^{(1)}$}^{-1}\right){c^u_{\left[\eta_i^{(1)}\right]_{(1,m)}}}^{\hspace{-5.5mm}\dagger\hspace{2mm}}c^v_{\left[\eta_j^{(1)}\right]_{(1,m)}},
    \label{eq:Ham-m-PBC}
\end{equation}
where we have set $\breve{\htgel}^{(m)}=1$, since we are now focusing on a single copy of the supercell $C^{(m)}$; however, we keep the summation over $\htgel^{(m)}\in\HTGsc{m}$ due to the possibility of long-range hopping processes in $\Ham$ that exceed the size of a single supercell.

Next, we introduce another level of subdivision: an even larger super-supercell $C^{(n)}$ with $n>m$ consisting of copies of the supercell $C^{(m)}$.
Using the unique decomposition given in \cref{eq:super-supercell-decomp}, we can rewrite \cref{eq:hopp-Ham:m-supercell} as
\begin{equation}
    \Ham = \sum_{\htgel^{(n)},\breve{\htgel}^{(n)}\in\HTGsc{n}}\sum_{\eta_{i'}^{(m)},\eta_{j'}^{(m)}\in\rtransv{\HTGsc{m}}{\HTGsc{n}}}\sum_{\eta_i^{(1)},\eta_j^{(1)}\in\rtransv{\HTGuc{}}{\HTGsc{m}}}\sum_{u,v} h^{uv}\left(\eta_i^{(1)}\eta_{i'}^{(m)}\htgel^{(n)}\mbox{$\breve{\htgel}^{(n)}$}^{-1}\mbox{$\eta_{j'}^{(m)}$}^{-1}\mbox{$\eta_j^{(1)}$}^{-1}\right){c^u_{\eta_i^{(1)}\eta_{i'}^{(m)}\htgel^{(n)}}}^{\hspace{-2.5mm}\dagger\hspace{2.5mm}}c^v_{\eta_j^{(1)}\eta_{j'}^{(m)}\breve{\htgel}^{(n)}},
    \label{eq:hopp-Ham:n-supercell}
\end{equation}
which once more can be restricted to a PBC cluster, in this case to $C^{(n)}$:
\begin{equation}
    \Ham_\text{$n$-PBC}  = \sum_{\htgel^{(n)}\in\HTGsc{n}}\sum_{\eta_{i'}^{(m)},\eta_{j'}^{(m)}\in\rtransv{\HTGsc{m}}{\HTGsc{n}}}\sum_{\eta_i^{(1)},\eta_j^{(1)}\in\rtransv{\HTGuc{}}{\HTGsc{m}}}\sum_{u,v} h^{uv}\left(\eta_i^{(1)}\eta_{i'}^{(m)}\htgel^{(n)}\mbox{$\eta_{j'}^{(m)}$}^{-1}\mbox{$\eta_j^{(1)}$}^{-1}\right){c^u_{\left[\eta_i^{(1)}\eta_{i'}^{(m)}\right]_{(1,n)}}}^{\hspace{-4.5mm}\dagger\hspace{3mm}}c^v_{\left[\eta_j^{(1)}\eta_{j'}^{(m)}\right]_{(1,n)}},
    \label{eq:Ham-n-PBC}
\end{equation}
where primitive cells are now labeled by cosets $[\eta_j^{(1)}\eta_{j'}^{(m)}]_{(1,n)}\in\HTGuc{}/\HTGsc{n}$ [note the difference with \cref{eq:Ham-m-PBC}].

For simplicity, we temporarily suppress the internal structure of $C^{(m)}$ in terms of copies of $C^{(1)}$, introducing new orbital labels $\mu=(u,\eta_i^{(1)})$, $\nu=(v,\eta_j^{(1)})$ and creation/annihilation operators
\begin{equation}
c^\mu_{\left[\eta_{i'}^{(m)}\right]_{(m,n)}} = c^u_{\left[\eta_i^{(1)}\eta_{i'}^{(m)}\right]_{(1,n)}} \qquad \textrm{and} \qquad c^\nu_{\left[\eta_{j'}^{(m)}\right]_{(m,n)}} = c^v_{\left[\eta_j^{(1)}\eta_{j'}^{(m)}\right]_{(1,n)}}, \label{eq:c-operators-shortened}  
\end{equation}
where $[\eta_{i'}^{(m)}]_{(m,n)}\in\HTGsc{m}/\HTGsc{n}$ now labels the copies of $C^{(m)}$, rather than copies of $C^{(1)}$ as did $[\eta_i^{(1)}\eta_{i'}^{(m)}]_{(1,n)}$, and $\mu$ takes care of copies of $C^{(1)}$ within $C^{(m)}$, such that still all copies of $C^{(1)}$ are captured.
Then,
\begin{equation}
    \Ham_\text{$n$-PBC} = \sum_{\htgel^{(n)}\in\HTGsc{n}}\sum_{\eta_{i'}^{(m)},\eta_{j'}^{(m)}\in\rtransv{\HTGsc{m}}{\HTGsc{n}}}\sum_{\mu,\nu} \HopMat^{\mu\nu}\left(\eta_{i'}^{(m)}\htgel^{(n)}\mbox{$\eta_{j'}^{(m)}$}^{-1}\right){c^\mu_{\left[\eta_{i'}^{(m)}\right]_{(m,n)}}}^{\hspace{-5.5mm}\dagger\hspace{3mm}}c^\nu_{\left[\eta_{j'}^{(m)}\right]_{(m,n)}},
    \label{eq:Ham-n-PBC:med}
\end{equation}
where
\begin{equation}
    \HopMat^{\mu\nu}\left(\eta_{i'}^{(m)}\htgel^{(n)}\mbox{$\eta_{j'}^{(m)}$}^{-1}\right)=h^{uv}\left(\eta_i^{(1)}\eta_{i'}^{(m)}\htgel^{(n)}\mbox{$\eta_{j'}^{(m)}$}^{-1}\mbox{$\eta_j^{(1)}$}^{-1}\right)
    \label{eq:HopMat-m-supercell}
\end{equation}
forms the hopping matrix $\HopMat$ defined on $C^{(m)}$.

\subsection{Bloch Hamiltonian}\label{Sec:HTB:Bloch-Ham}

In this subsection, we derive the Bloch Hamiltonian corresponding to the most general $\HTGuc{}$-translation-invariant hopping Hamiltonian.
As discussed in the main text, the Bloch Hamiltonian depends on the choice of unit cell, i.e., the choice of translation subgroup that is considered.
We assume a subdivision of the infinite lattice into supercells $C^{(m)}$ and ignore the additional internal translation-invariant structure of $C^{(m)}$ in terms of primitive cells, i.e., only $\HTGsc{m}$ is utilized.
However, instead of doing that for the Hamiltonian defined on the infinite lattice [\cref{eq:hopp-Ham:m-supercell}], we assume a finite but large PBC cluster made up of many copies of $C^{(m)}$.
More precisely, we choose this finite PBC cluster to be a larger super-supercell $C^{(n)}$, i.e., we consider \cref{eq:Ham-n-PBC:med}.
This reduces the full translation group $\HTGsc{m}$ to the quotient group $\HTGsc{m}/\HTGsc{n}$, such that we avoid dealing with the former infinite group.
Below, we first show that the Hamiltonian given in \cref{eq:Ham-n-PBC:med} can be block-diagonalized into blocks of Bloch Hamiltonians [\cref{eq:Ham-n-PBC:Bloch-decomp,eq:Ham-n-PBC:Bloch-Ham}].

Only at the end do we take the limit $n\to\infty$, thereby recovering the infinite lattice~\cite{Lux:2022}.
This corresponds to taking the limit of an infinitely large super-supercell $C^{(n)}$.
In the main text, we have discussed the reciprocal space of the infinite lattice, which is the continuous space of irreducible representations (IRs) of the infinite group $\HTGsc{m}$.
Alternatively, we could consider the discrete reciprocal space of a finite PBC cluster, i.e., the super-supercell $C^{(n)}$.
This reciprocal space is spanned by the finitely many IRs of the \emph{finite} group $\HTGsc{m}/\HTGsc{n}$~\cite{Maciejko:2022}.
By the third isomorphism theorem~\cite{Robinson:1996}, the sequence of normal subgroups $\HTGsc{m+1}\nsubgstr\HTGsc{m}$ implies that, for fixed $n$, the quotients $\HTGsc{m}/\HTGsc{n}$ also form a sequence of normal subgroups:
\begin{equation}
    \HTGsc{m+1}/\HTGsc{n}\nsubgstr \HTGsc{m}/\HTGsc{n}.
\end{equation}
Therefore, the subduction and induction of representations apply to that space as well.
We anticipate that this provides an approximation of the thermodynamic limit~\cite{Lux:2022} and converges to the full reciprocal space for $n\to\infty$.

Starting from \cref{eq:Ham-n-PBC:med} we now derive the corresponding Bloch Hamiltonian implementing the translation symmetry due to the finite translation group $\HTGsc{m}/\HTGsc{n}$.
Simplifying the notation, namely writing $\tilde{\htgel}$ for $\htgel^{(n)}$ and dropping the super- and subscripts of $\eta_i^{(m)}$ and of the cosets $[\cdot]_{(m,n)}$, respectively, the Hamiltonian on the finite $C^{(n)}$ PBC cluster reads
\begin{equation}
    \Ham_\text{$n$-PBC} = \sum_{\tilde{\htgel}\in\HTGsc{n}}\sum_{\eta_i,\eta_j\in\rtransv{\HTGsc{m}}{\HTGsc{n}}}\sum_{\mu,\nu} \HopMat^{\mu\nu}\left(\eta_i\tilde{\htgel}\eta_j^{-1}\right){c^\mu_{\left[\eta_i\right]}}^{\hspace{-1mm}\dagger\hspace{1mm}}c^\nu_{\left[\eta_j\right]}.
\end{equation}
We recognize that $\eta_i\tilde{\htgel} = \htgel'\in\HTGsc{m}$ and by the coset decomposition of $\HTGsc{m}/\HTGsc{n}$ the sums over $\tilde{\htgel}\in\HTGsc{n}$ and $\eta_i\in\rtransv{\HTGsc{m}}{\HTGsc{n}}$ can be replaced by a sum over $\htgel'\in\HTGsc{m}$:
\begin{equation}
    \Ham_\text{$n$-PBC} = \sum_{\htgel'\in\HTGsc{m}}\sum_{\eta_j\in\rtransv{\HTGsc{m}}{\HTGsc{n}}}\sum_{\mu,\nu} \HopMat^{\mu\nu}\left(\htgel'\eta_j^{-1}\right){c^\mu_{\left[\htgel'\right]}}^{\hspace{-1mm}\dagger\hspace{1mm}}c^\nu_{\left[\eta_j\right]},
\end{equation}
where in the subscript of the creation operator we used that $\eta_i$ and $\htgel'=\eta_i\tilde{\htgel}$ are in the same coset of $\HTGsc{m}/\HTGsc{n}$.
Next, for a given $\eta_j\in\rtransv{\HTGsc{m}}{\HTGsc{n}}$, $\htgel'$ can be uniquely written as $\htgel\eta_j$ with $\htgel=\htgel'\eta_j^{-1}\in\HTGsc{m}$, allowing us to introduce the sum over $\htgel\in\HTGsc{m}$ instead:
\begin{equation}
    \Ham_\text{$n$-PBC} = \sum_{\eta_j\in\rtransv{\HTGuc{}}{\HTGsc{m}}}\sum_{\htgel\in\HTGsc{m}}\sum_{\mu,\nu} \HopMat^{\mu\nu}\left(\htgel\right){c^\mu_{\left[\htgel\eta_j\right]}}^{\hspace{-2mm}\dagger\hspace{2mm}}c^\nu_{\left[\eta_j\right]}.
    \label{eq:Ham-n-PBC:short}
\end{equation}

According to Eqs.~(25-26) in Ref.~\onlinecite{Maciejko:2022},
\begin{equation}
    {c^\mu_{[\htgel\eta_j]}}^{\hspace{-2mm}\dagger\hspace{2mm}} = \sum_i {c^\mu_{[\eta_i]}}^{\hspace{-1.5mm}\dagger\hspace{1.5mm}}\mathcal{U}_{ij}\left(\left[\htgel\right]\right),
\end{equation}
where the matrices
\begin{equation}
    \mathcal{U}_{ij}\left(\left[\htgel\right]\right) = \delta_{[\eta_i],[\htgel\eta_j]}
\end{equation}
form the so-called \emph{regular representation} of $\HTGsc{m}/\HTGsc{n}$ (note that in contrast to Ref.~\onlinecite{Maciejko:2022}, we are using right cosets).
The regular representation can be decomposed into a direct sum of all IRs, each appearing with multiplicity equal to their dimension $d_\lambda$~\cite{Maciejko:2022}, i.e., there exists a unitary transformation $P$ such that
\begin{equation}
    P\mathcal{U}\left(\left[\htgel\right]\right)P^{-1} = \bigoplus_{\lambda=1}^{\mathcal{N}}d_\lambda D^{(\lambda)}\left(\left[\htgel\right]\right),
\end{equation}
where $\mathcal{N}$ is the number of conjugacy classes of $\HTGsc{m}/\HTGsc{n}$.

The above implies that $\Ham_\text{$n$-PBC}$ can be written in terms of those blocks:
\begin{align}
    \Ham_\text{$n$-PBC} &= \sum_{\htgel\in\HTGsc{m}}\sum_{i,j}\sum_{\mu,\nu} \HopMat^{\mu\nu}\left(\htgel\right){c^\mu_{\left[\eta_i\right]}}^{\hspace{-1.5mm}\dagger\hspace{1.5mm}}\mathcal{U}_{ij}\left(\left[\htgel\right]\right)c^\nu_{\left[\eta_j\right]}\\
    &= \sum_{k,l}\sum_{\mu,\nu}\adjo{\left(\sum_{i}P_{ki}c^\mu_{\left[\eta_{i}\right]}\right)}\sum_{\htgel\in\HTGsc{m}} \HopMat^{\mu\nu}\left(\htgel\right)\left(\bigoplus_{\lambda=1}^{\mathcal{N}}d_\lambda D^{(\lambda)}\left(\left[\htgel\right]\right)\right)_{kl}\sum_{j}P_{lj}c^\nu_{\left[\eta_{j}\right]}\\
    &= \sum_{k,l}\sum_{\mu,\nu}\adjo{\left(\sum_{i}P_{ki}c^\mu_{\left[\eta_{i}\right]}\right)}\left(\bigoplus_{\lambda=1}^{\mathcal{N}}d_\lambda \sum_{\htgel\in\HTGsc{m}} \HopMat\left(\htgel\right)\otimes D^{(\lambda)}\left(\left[\htgel\right]\right)\right)^{\mu\nu}_{kl}\sum_{j}P_{lj}c^\nu_{\left[\eta_{j}\right]},
    \label{eq:Ham-n-PBC:Bloch-decomp}
\end{align}
where we recognize the Bloch Hamiltonian
\begin{equation}
    \BlochHam(D) = \sum_{\htgel\in\HTGsc{m}}\HopMat(\htgel)\otimes D\left(\left[\htgel\right]\right)
    \label{eq:Ham-n-PBC:Bloch-Ham}
\end{equation}
with the hopping matrix $\HopMat(\htgel)$ on the supercell $C^{(m)}$ having components $\HopMat^{\mu\nu}(\htgel)$.

Recovering the subdivision into primitive cells using \cref{eq:c-operators-shortened}, we have
\begin{align}
    \Ham_\text{$n$-PBC} &= \sum_{k,l}\sum_{i,j}\sum_{u,v}\adjo{\mbox{$\hat{c}^u_{i,k}$}}\sum_{\htgel^{(m)}\in\HTGsc{m}} \underbrace{h^{uv}\left(\eta_i^{(1)}\htgel^{(m)}\mbox{$\eta_j^{(1)}$}^{-1}\right)}_{=:\HopMat^{uv}_{ij}\left(\htgel^{(m)}\right)}\left(\bigoplus_{\lambda=1}^{\mathcal{N}}d_\lambda D^{(\lambda)}\left(\left[\htgel^{(m)}\right]_{(m,n)}\right)\right)_{kl}\hat{c}^v_{j,l}\\
    &= \sum_{k,l}\sum_{i,j}\sum_{u,v}\adjo{\mbox{$\hat{c}^u_{i,k}$}}\left(\bigoplus_{\lambda=1}^{\mathcal{N}}d_\lambda\BlochHam\left(D^{(\lambda)}\right)\right)_{ijkl}^{uv}\hat{c}^v_{j,l}
    \label{eq:Ham-n-PBC-full:Bloch-decomp}
\end{align}
with
\begin{equation}
    \hat{c}^u_{i,k} = \sum_{i'} P_{ki'} c^u_{\left[\eta_i^{(1)}\eta_{i'}^{(m)}\right]_{(1,n)}},
    \label{eq:Ham-n-PBC-full:Bloch-c-operators}
\end{equation}
and the Bloch Hamiltonian $\BlochHam\left(D\right)$ with matrix elements
\begin{equation}
    \BlochHam_{ijk^{(\lambda)} l^{(\lambda)}}^{uv}\left(D^{(\lambda)}\right) = \sum_{\htgel^{(m)}\in\HTGsc{m}}\HopMat_{ij}^{uv}\left(\htgel^{(m)}\right)D^{(\lambda)}_{k^{(\lambda)}l^{(\lambda)}}\left(\left[\htgel^{(m)}\right]_{(m,n)}\right).
    \label{eq:Ham-n-PBC-full:Bloch-Ham}
\end{equation}
To avoid confusion, let us at this point remind ourselves of the meaning of all the indices appearing in \cref{eq:Ham-n-PBC-full:Bloch-decomp,eq:Ham-n-PBC-full:Bloch-c-operators,eq:Ham-n-PBC-full:Bloch-Ham}.
The superscript $(m)$ labels the translation group $\HTGsc{m}$ of the supercell $C^{(m)}$ and the subscript $(m,n)$ indicates that $[\htgel^{(m)}]_{(m,n)}$ denotes the coset $\htgel^{(m)}\HTGsc{n}$.
The IRs $D^{(\lambda)}$ of the finite group $\HTGsc{m}/\HTGsc{n}$ are labeled by $\lambda$ and they span the reciprocal space of the PBC clusters formed by the super-supercell $C^{(n)}$.
For a specific IR $D^{(\lambda)}$ the matrix indices $k^{(\lambda)}$,$l^{(\lambda)}$ range over the dimensions of the IR (and never arise in the Euclidean Bloch Hamiltonian due to the fact that Euclidean translation groups only have 1D IRs).
The subscripts $i,j$ label the copies of the primitive cell $C^{(1)}$ in the supercell $C^{(m)}$ and therefore range from $1$ to $|\HTGuc{}/\HTGsc{m}|$, while $u,v$ range over the sites in the primitive cell $C^{(1)}$ (and therefore do not arise if applying Bloch's theorem to a single primitive cell).

Finally, we take the limit $n\to\infty$, i.e., we let the super-supercell $C^{(n)}$ encompass the whole infinite lattice.
Formally, this implies the replacement of $\HTGsc{m}/\HTGsc{n}$ by $\HTGsc{m}$, such that \cref{eq:hopp-Ham:m-supercell} is block-diagonalized as
\begin{align}
    \Ham &= \sum_{k,l}\sum_{i,j}\sum_{u,v}\adjo{\mbox{$\hat{c}^u_{i,k}$}}\sum_{\htgel^{(m)}\in\HTGsc{m}} \HopMat^{uv}_{ij}\left(\htgel^{(m)}\right)\left(\bigoplus_{\lambda} d_\lambda D^{(\lambda)}\right)_{kl}\left(\htgel^{(m)}\right)\hat{c}^v_{j,l}\\
    &= \sum_{k,l}\sum_{i,j}\sum_{u,v}\adjo{\mbox{$\hat{c}^u_{i,k}$}}\left(\bigoplus_{\lambda} d_\lambda\BlochHam\left(D^{(\lambda)}\right)\right)_{ijkl}^{uv}\hat{c}^v_{j,l}
    \label{eq:Bloch-decomposition:infinite}
\end{align}
with Bloch Hamiltonian given in \cref{main:eq:Bloch-Hamiltonian} of the main text,
\begin{equation}
    \BlochHam(D) = \sum_{\htgel^{(m)}\in\HTGsc{m}}\HopMat\left(\htgel^{(m)}\right)\otimes D\left(\htgel^{(m)}\right),
    \label{eq:Bloch-Hamiltonian}
\end{equation}
where $D$ is now an IR of $\HTGsc{m}$, and
\begin{equation}
    \hat{c}^u_{i,k} = \sum_{i'} P_{ki'} c^u_{\eta_i^{(1)}\eta_{i'}^{(m)}}.
\end{equation}
Let us remark that in \cref{eq:Bloch-decomposition:infinite}, we have kept the symbol \enquote{$\oplus$}, even though in reality the direct sum now goes over a \emph{continuous} space of infinitely many IRs and a suitable measure over the representation space needs to be introduced~\cite{Nagy:2022}.
We leave the question of a mathematical formalization of the $n\to\infty$ limit for future studies.

In this work, we only ever \emph{explicitly} sample the subspace of one-dimensional (1D) IRs, i.e., the Abelian Brillouin zone (ABZ), which for the supercell $C^{(m)}$ is given by
\begin{equation}
    \ABZ{m} = \left\{\htg{i}^{(m)}\mapsto D_{\vec{k}^{(m)}}\left(\htg{i}^{(m)}\right) = \e^{\i k_i^{(m)}}\,:\,\vec{k}^{(m)}\in \torus{2\genus{}^{(m)}}\right\},
    \label{eq:ABZ}
\end{equation}
with the generators $\htg{i}^{(m)}$ of $\HTGsc{m}$.
The \emph{Abelian Bloch Hamiltonian} is then given according to \cref{eq:Bloch-Hamiltonian}:
\begin{equation}
    \BlochHam^{(m)}\left(\vec{k}^{(m)}\right) = \sum_{\htgel^{(m)}\in\HTGsc{m}} \HopMat\left(\htgel^{(m)}\right)\e^{\i\sum_{i=1}^{2\genus{}^{(m)}}K_ik_i},
    \label{eq:Abelian-Bloch-Hamiltonian}
\end{equation}
where $\genus{}^{(m)}$ is the genus of the compactified $C^{(m)}$ cell, and $K_i$ the number of times the generator $\htgel_i^{(m)}$ appears in $\htgel^{(m)}$ minus the number of times its inverse appears.

\section{Hopping models}

In this section, we give additional details on the models discussed in the main text.
We specifically:
\begin{enumerate}[(i)]
    \item provide explicit expressions for the Hamiltonians and for the quotient groups $\TG/\HTGsc{m}$ specifying the supercells in \cref{Sec:tessNN-models,Sec:kagomeNN-models,Sec:Haldane-model,Sec:BBH-model}, 
    \item illustrate the primitive cells and the model definitions of the Hamiltonians in \cref{fig:models} and illustrate their supercells in \cref{fig:supercell-sequences},
    \item show their density of states (DOS) in more detail in \cref{fig:DOS:NN-models,fig:DOS:others}, and give the necessary parameters required to reproduce the data, i.e., sampling parameters.
\end{enumerate}
Note that in the supplementary data and code~\cite{SDC} we make available the graphs representing the triangular tessellations, the lattices, the model Hamiltonians, and the supercells produced with our software package~\cite{HyperCells,HyperBloch}, as well as the DOS data.
Furthermore, we provide example code on how to obtain and handle those objects.

\subsection{Nearest-neighbor hopping models on \texorpdfstring{$\{p,q\}$}{\{p,q\}} lattices}\label{Sec:tessNN-models}

We start with the nearest-neighbor (NN) hopping models on $\{p,q\}$ lattices, whose sites are given by only the elements of $V_y\cong\rtransv{\TG^+}{\TG_y^+}$, i.e., the $q$-fold rotation-symmetric vertices of the Schwarz triangles of $\TG(2,q,p)$.
In the main text, we have only briefly discussed the $\{8,8\}$ lattice, for which we have not shown all the data.
Here, we show the full data in \cref{Sec:88-tess-NN} and \cref{fig:DOS:NN-models:88}.
While the NN hopping model on the $\{8,3\}$ lattice was not explicitly discussed in the main text, we briefly discuss it here because of its relation to the NN hopping model on the octagon-kagome lattice and to the $\{8,3\}$-Haldane model, which all share $\TG(2,3,8)$ as their space group.

\subsubsection{\texorpdfstring{$\{8,8\}$}{\{8,8\}} lattice}\label{Sec:88-tess-NN}
\begin{figure}[p]
    \centering
    \subfloat{\label{fig:models:88-NN}}
    \subfloat{\label{fig:models:83-NN}}
    \subfloat{\label{fig:models:8kagome-NN}}
    \subfloat{\label{fig:models:83-Haldane}}
    \includegraphics{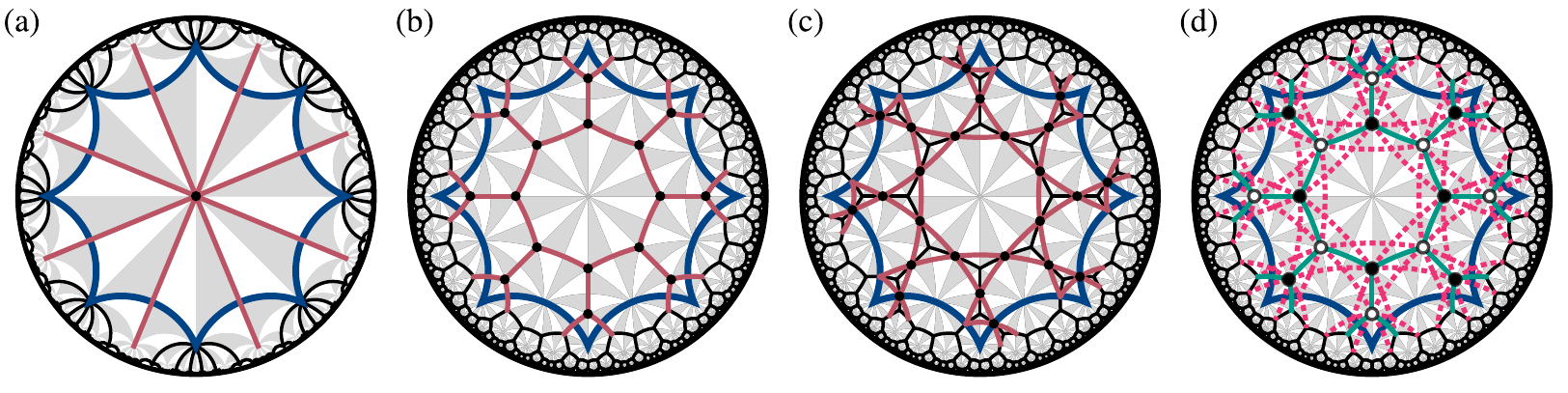}
    \caption{Model definitions: primitive cells (blue polygon) with sites (black dots) and hoppings (red lines) for (a) the nearest-neighbor (NN) model on the $\{8,8\}$ lattice, (b) the NN model on the $\{8,3\}$ lattice, (c) the NN model on the octagon-kagome lattice (the line graph of the $\{8,3\}$ lattice, and (d) the Haldane model on the $\{8,3\}$ lattice).
    (d) The sublattice potential $\pm h_0$ is indicated by filled and empty disks on the sites, the NN hoppings $h_1$ are shown by green solid lines and the next-NN hoppings with amplitudes $h_2\e^{\pm\i\phi}$ by dashed magenta lines.}
    \label{fig:models}
\end{figure}

The $\{8,8\}$ lattice is defined by the hyperbolic tessellation where eight octagons meet at each vertex, cf.~\cref{fig:models:88-NN}.
Its space group is the triangle group $\TG(2,8,8)$ and the minimal unit cell, i.e., the primitive cell (depicted as a blue polygon in \cref{fig:models:88-NN}) is given by the smallest quotient $\tgquot{2}{6}$ of $\TG$ by one of its normal subgroups $\HTGuc{}$~\cite{Conder:2007}:
\begin{equation}
    \HPGuc{} = \gpres{a,b,c}{a^2,b^2,c^2,(ab)^2,(bc)^8,(ca)^8,(abc)^2,(bc)^3(ca)^{-1}},
\end{equation}
with the generators $a$, $b$, $c$ the reflections across the sides of the Schwarz triangle, cf.~\cref{main:fig:triangle_group}.
With $\HPGuc{}$ isomorphic to the quotient group $\TG/\HTGuc{}$, $\HTGuc{}$ is given by the kernel of that isomorphism, which can be extracted using \gap{}.
With our algorithm for constructing symmetric unit cells~\cite{HyperCells}, we obtain
\begin{align}
    \HTGuc{}  &= \gpres{\htg{1},\htg{2},\htg{3},\htg{4}}{\htg{2}\htg{1}^{-1}\htg{4}^{-1}\htg{3}\htg{2}^{-1}\htg{1}\htg{4}\htg{3}^{-1}}\\
    &= \gpres{\htg{1},\htg{2},\htg{3},\htg{4}}{\htg{1}\htg{2}^{-1}\htg{3}\htg{4}^{-1}\htg{1}^{-1}\htg{2}\htg{3}^{-1}\htg{4}}
\end{align}
and recognize the translation group of the Bolza cell which has been discussed extensively in recent literature~\cite{Maciejko:2021,Maciejko:2022,Urwyler:2022,Chen:2023} on hyperbolic tight-binding models, since it is the maximal translation group of the $\{8,8\}$, $\{8,3\}$, and $\{8,4\}$ lattices~\cite{Boettcher:2022}.

\begin{figure}[p]
    \centering
    \subfloat{\label{fig:supercell-sequences:288}}
    \subfloat{\label{fig:supercell-sequences:238}}
    \subfloat{\label{fig:supercell-sequences:246}}
    \includegraphics{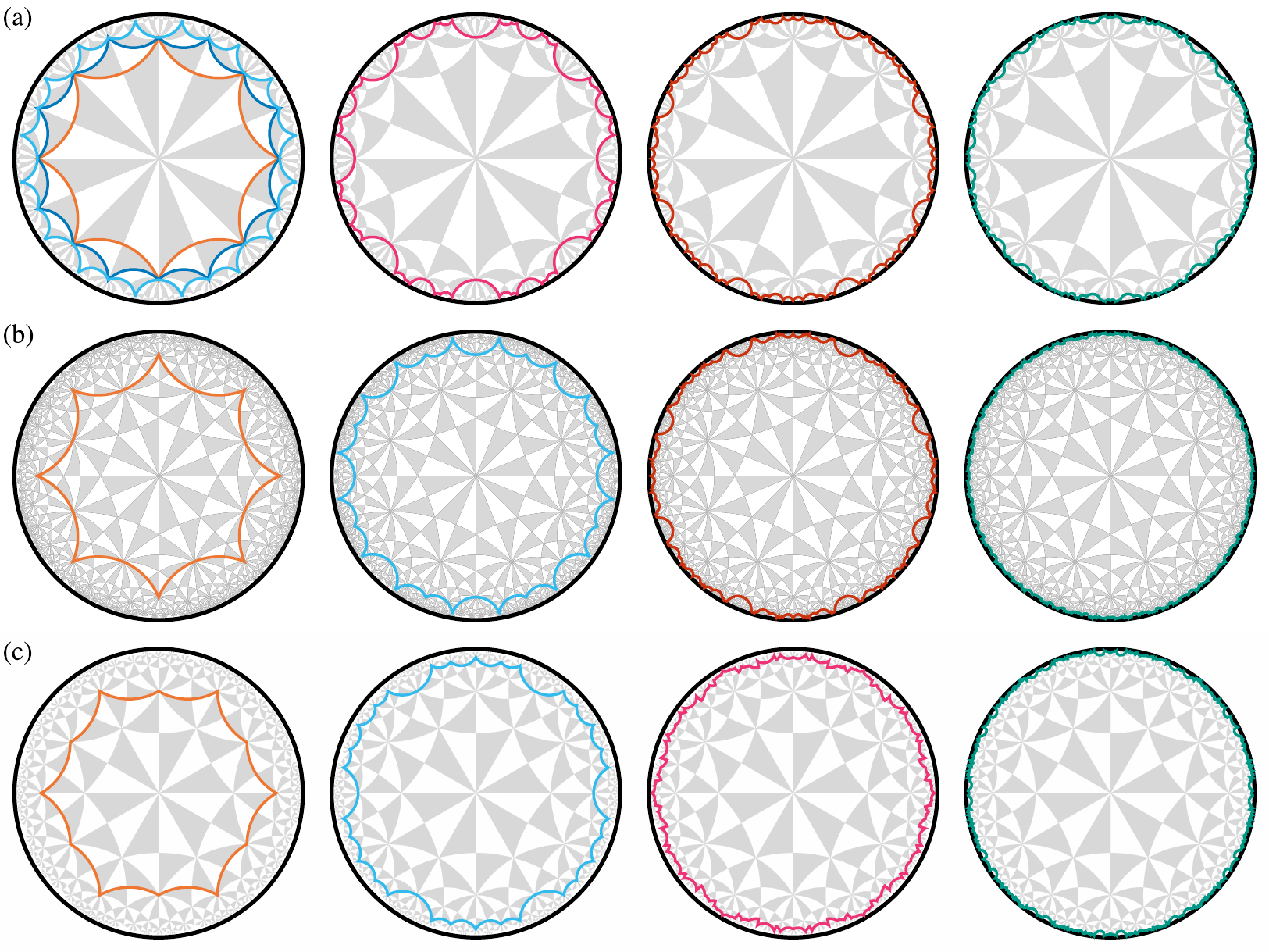}
    \caption{
        Supercell sequences used to compute the density of states shown in the main text and in \cref{fig:DOS:NN-models,fig:DOS:others}. The supercells are specified by the corresponding quotients $\HPG\cong\TG/\HTG$ in the form ``$T\genus{}.j$'', denoting the $j^\mathrm{th}$ quotient of any triangle group $\TG(r,q,p)$ having genus $\genus{}$~\cite{Conder:2007}.
        Note that all the sequences start with a genus-$2$ unit cell, such that, per the Riemann-Hurwitz formula, $N=\genus{}-1$ is the number of unit cells in the supercell.
        (a) $\TG(2,8,8)$, relevant for the $\{8,8\}$ lattice: $\tgquot{2}{6}$, $\tgquot{3}{11}$, $\tgquot{5}{13}$, $\tgquot{9}{20}$, $\tgquot{17}{29}$, $\tgquot{33}{44}$, [$\tgquot{65}{78}$ (not shown)].
        (b) $\TG(2,3,8)$, relevant for the $\{8,3\}$ lattice and the octagon-kagome lattice: $\tgquot{2}{1}$, $\tgquot{5}{1}$, $\tgquot{17}{2}$, $\tgquot{33}{1}$. (Note that the first two of the constructed super-cells match precisely those found for translation groups $\tgquot{2}{6}$ and $\tgquot{5}{13}$ for the $\Delta(2,8,8)$ lattice.)
        (c) $\TG(2,4,6)$, relevant for the $\{6,4\}$ lattice: $\tgquot{2}{2}$, $\tgquot{5}{4}$, $\tgquot{9}{3}$, $\tgquot{33}{11}$, [$\tgquot{65}{9}$ (not shown)].
    }
    \label{fig:supercell-sequences}
\end{figure}

Using \gap{}, we have found a normal sequence of translation subgroups $\HTGsc{m}\nsubgstr\TG(2,8,8)$ via the appropriate quotient groups $\HPG^{(m)}\cong\TG/\HTGsc{m}$ given in Ref.~\onlinecite{Conder:2007}, whose supercells preserve a maximum amount of symmetry.
Using
\begin{equation}
    x = ab,\quad y = bc,\quad z = ca,
    \label{eq:rotation-elements}
\end{equation}
the quotient groups have the following presentations:
\begin{equation}
  \begin{alignedat}{3}
    \HPG^{(1)} &= \HPG^{\tgquot{2}{6}} &&= \gpres{a,b,c}{a^2,b^2,c^2,x^2,y^8,z^8,xzy,y^3z^{-1}},\\
    \HPG^{(2)} &= \HPG^{\tgquot{3}{11}} &&= \gpres{a,b,c}{a^2,b^2,c^2,x^2,y^8,z^8,xzy},\\
    \HPG^{(3)} &= \HPG^{\tgquot{5}{13}} &&= \gpres{a,b,c}{a^2,b^2,c^2,x^2,y^8,z^8,xy^{-2}z^{-1}y,xzy^{-1}z^{-2}},\\
    \HPG^{(4)} &= \HPG^{\tgquot{9}{20}} &&= \gpres{a,b,c}{a^2,b^2,c^2,x^2,y^8,z^8,xy^{-2}z^{-1}y},\\
    \HPG^{(5)} &= \HPG^{\tgquot{17}{29}} &&= \gpres{a,b,c}{a^2,b^2,c^2,x^2,y^8,z^8,xzy^{-2}z^{-2}y,(yz^{-1}y^2)^2},\\
    \HPG^{(6)} &= \HPG^{\tgquot{33}{44}} &&= \gpres{a,b,c}{a^2,b^2,c^2,x^2,y^8,z^8,xzy^{-2}z^{-2}y},\\
    \HPG^{(7)} &= \HPG^{\tgquot{65}{78}} &&= \gpres{a,b,c}{a^2,b^2,c^2,x^2,y^8,z^8, xy^{-1}zy^{-2}(z^{-1}y)^2, x(zy^{-1})^2z^{-2}yz^{-1}, yxz^2y^2xzy^{-1}z^{-1}}.
  \end{alignedat}
  \label{eq:supercell-sequence:288}
\end{equation}
The corresponding connected symmetric supercells are illustrated in \cref{fig:supercell-sequences:288} and their triangle group graphs as well as boundary identifications are given in the supplementary data and code~\cite{SDC}.
Note that by construction they all preserve the symmetries that leave their center invariant, i.e., eight-fold rotation and mirrors through the center.

Each octagon of the $\{8,8\}$ tessellation corresponds to a copy of the primitive unit cell and there is a single site in its center shown as a black dot in \cref{fig:models:88-NN}.
The NN model on the $\{8,8\}$ lattice thus has the Hamiltonian
\begin{equation}
    \Ham = -\sum_{\nn{i,j}}\adjo{c}_ic_j,
\end{equation}
where $\nn{i,j}$ denotes the centers of two neighboring octagons.
The hopping processes starting in the primitive unit cell are illustrated as red lines in \cref{fig:models:88-NN} and the real-space hopping Hamiltonian is given in graph form in the supplementary data and code~\cite{SDC}.

\begin{figure}
    \centering
    \subfloat{\label{fig:DOS:NN-models:88}}
    \subfloat{\label{fig:DOS:NN-models:83}}
    \includegraphics{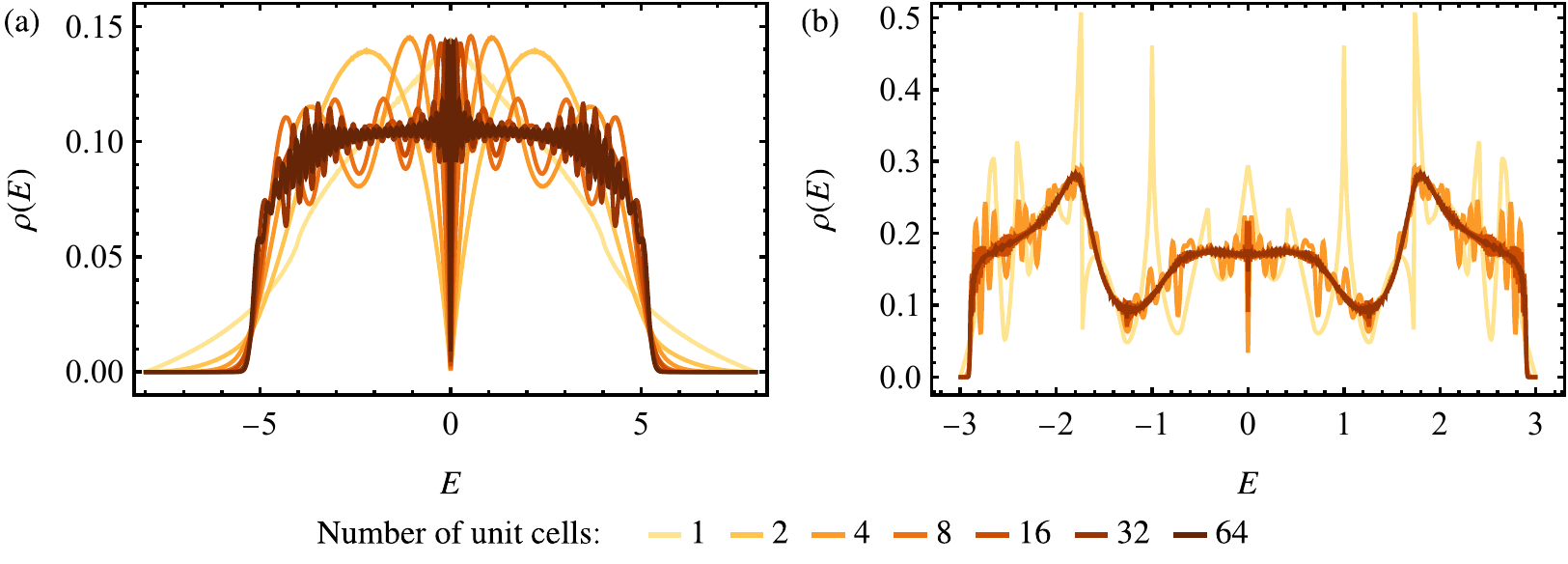}
    \caption{
        Density of states of the nearest-neighbor hopping model on (a) the $\{8,8\}$ and (b) the $\{8,3\}$ lattice obtained using the supercell method.
        In (a) the supercell sequence $\tgquot{2}{6}$, $\tgquot{3}{11}$, $\tgquot{5}{13}$, $\tgquot{9}{20}$, $\tgquot{17}{29}$, $\tgquot{33}{44}$, $\tgquot{65}{78}$ with $1$, $2$, $4$, $8$, $16$, $32$, $64$ unit cells per supercell was used, while in (b) the sequence $\tgquot{2}{1}$, $\tgquot{5}{1}$, $\tgquot{17}{2}$, $\tgquot{33}{1}$ with $1$, $4$, $16$, $32$ unit cells was used.
        In both cases, the energy resolution is $0.005$ with a moving average with window $0.01$.
    }
    \label{fig:DOS:NN-models}
\end{figure}

For each cell $C^{(m)}$ in the sequence given by \cref{eq:supercell-sequence:288}, we diagonalized the Abelian Bloch Hamiltonian $\BlochHam^{(m)}\left(\vec{k}^{(m)}\right)$ [given in \cref{eq:Abelian-Bloch-Hamiltonian}] for $10^9$ randomly sampled points in $\torus{2\genus{}^{(m)}}$ [corresponding to the ABZ given in \cref{eq:ABZ}].
From the eigenenergies thus obtained, we computed the DOS $\rho^{(m)}(E)$ for an energy resolution of $\dd{E}=0.005$, which was then smoothed using a moving average with window $2\dd{E}=0.01$, cf.~\cref{fig:DOS:NN-models:88}.
In \cref{main:fig:supercell:dos88} in the main text we additionally show the additional red curve, which was obtained by extrapolating the data in \cref{fig:DOS:NN-models:88} by eye, observing the fact that the DOS near the band edges becomes sharper, while the oscillations on the plateau are becoming smaller for larger supercells.
We provide the raw data for the DOS in the supplementary data and code~\cite{SDC}.

\subsubsection{\texorpdfstring{$\{8,3\}$}{\{8,3\}} lattice}
The $\{8,3\}$ lattice is defined by the hyperbolic tessellation where three octagons meet at each vertex, cf.~\cref{fig:models:83-NN}, and can therefore be considered a generalization of the honeycomb lattice~\cite{Urwyler:2022}.
Its space group is the triangle group $\TG(2,3,8)$ and the primitive cell (blue polygon in \cref{fig:models:83-NN}) is given by the quotient group $\TG/\HTGuc{}$ denoted by $\tgquot{2}{1}$~\cite{Conder:2007}:
\begin{equation}
    \HPGuc{} = \gpres{a,b,c}{a^2,b^2,c^2,(ab)^2,(bc)^3,(ca)^8,c(abc)^2a(abc)^{-1}c^{-1}abca},
\end{equation}
resulting in the translation group
\begin{align}
    \HTGuc{} &= \gpres{\htg{1},\htg{2},\htg{3},\htg{4}}{\htg{4}\htg{1}\htg{2}^{-1}\htg{3}\htg{4}^{-1}\htg{1}^{-1}\htg{2}\htg{3}^{-1}}\\
    &= \gpres{\htg{1},\htg{2},\htg{3},\htg{4}}{\htg{1}\htg{2}^{-1}\htg{3}\htg{4}^{-1}\htg{1}^{-1}\htg{2}\htg{3}^{-1}\htg{4}},
\end{align}
which we once more identify as the translation group of the Bolza cell.

Using \gap{}, we have found a normal sequence of translation subgroups $\HTGsc{m}\nsubgstr\TG(2,3,8)$ via the appropriate quotient groups $\HPG^{(m)}\cong\TG/\HTGsc{m}$ given in Ref.~\onlinecite{Conder:2007}: With $x$, $y$, and $z$ given in \cref{eq:rotation-elements},
\begin{equation}
  \begin{alignedat}{3}
    \HPG^{(1)} &= \HPG^{\tgquot{2}{1}} &&= \gpres{a,b,c}{a^2,b^2,c^2,x^2,y^3,z^8,zyxz(zy)^{-1}xz},\\
    \HPG^{(2)} &= \HPG^{\tgquot{5}{1}} &&= \gpres{a,b,c}{a^2,b^2,c^2,x^2,y^3,z^8,z^3yz^{-1}xzy^{-1}xy^{-1}z^{-2}x},\\
    \HPG^{(3)} &= \HPG^{\tgquot{17}{2}} &&= \gpres{a,b,c}{a^2,b^2,c^2,x^2,y^3,z^8,(z^2yx)^2(zy^{-1}z^{-1}x)^2},\\
    \HPG^{(4)} &= \HPG^{\tgquot{33}{1}} &&= \gpres{a,b,c}{a^2,b^2,c^2,x^2,y^3,z^8,xz^2(zyx)^3z^2y^2z^{-2}xy^{-1}z^{-2}},
  \end{alignedat}
  \label{eq:supercell-sequence:238}
\end{equation}
with the corresponding connected and symmetric supercells illustrated in \cref{fig:supercell-sequences:238} and  their triangle group graphs as well as boundary identifications given in the supplementary data and code~\cite{SDC}.

The $\{8,3\}$ lattice is formed by all the three-fold symmetric points and the NN model has the Hamiltonian
\begin{equation}
    \Ham = -\sum_{\nn{i,j}}\adjo{c}_ic_j,
\end{equation}
where $\nn{i,j}$ denotes the hopping along a side of an octagon, cf.~\cref{fig:models:83-NN}.
The real-space hopping Hamiltonian is given in graph form in the supplementary data and code~\cite{SDC}.
Again we consider the ABZ and Bloch Hamiltonian given in \cref{eq:ABZ,eq:Abelian-Bloch-Hamiltonian}, respectively, but now for the sequence of supercells given in \cref{eq:supercell-sequence:238}.
We randomly sampled $10^8$ points in $\torus{2\genus{}^{(m)}}$, computed the spectrum of $\BlochHam^{(m)}\left(\vec{k}^{(m)}\right)$ and from that the DOS $\rho^{(m)}(E)$ for an energy resolution of $\dd{E}=0.005$, which was then smoothed using a moving average with window $2\dd{E}=0.01$, cf.~\cref{fig:DOS:NN-models:83}.
We provide the raw data for the DOS in the supplementary data and code~\cite{SDC}.

\subsubsection{Origin of the observed dips in the computed spectra at \texorpdfstring{$E=0$}{E=0}}

One systematic feature appearing in the presented DOS curves for the elementary NN model on both the $\{8,3\}$ and the $\{8,8\}$ lattice is the sharp dip at $E=0$. 
This feature becomes narrower for computations on larger supercells, and is expected to vanish altogether when extrapolating to the thermodynamic limit. 
Below, we trace the appearance of this undesired behavior back to a chiral symmetry of these particular Hamiltonians, and explain how this feature is removed when enlarging the supercell. 
For concreteness, we explicitly consider the $\{8,8\}$ model although the presented arguments can be adapted for any NN model on a bipartite lattice. 

The elementary nearest-neighbor model on the $\{8,8\}$ lattice has a single orbital per primitive cell, such that the Abelian Bloch states on that primitive cell form a single continuous band:
\begin{equation}
    E(\vec{k}) = -2\sum_{i=1}^4\cos(k_i),
\end{equation}
in the 4D Abelian Brillouin zone (BZ).
The condition $E=0$ implies that the states at zero energy form a (Fermi) surface with codimension $1$ (i.e., a 3D hypersurface), resulting in a finite DOS at $E=0$.
    
When considering a supercell, e.g., the $2$-supercell, two things happen.
First, the 4D BZ is reduced and the bands are folded, resulting in two bands in the reduced BZ. 
The specifics of this particular BZ reduction induced by extending the primitive cell to the $2$-supercell are discussed in the Supplementary Material of Ref.~\onlinecite{Tummuru:2023}.
The two obtained bands trivially intersect at $E=0$; yet, the computed DOS is unaffected by this step, because the band folding does not alter the collection of the captured eigenstates. 
In particular, the two bands are not coupled in the reduced BZ. 
This is because states in both bands still respect the full periodicity in primitive cells and carry a different value of a quantum number, namely of the initial four-component momentum.
    
Second, and this is the special feature of hyperbolic lattices, two more momentum components are introduced, increasing the BZ dimension from $4$ to $6$. 
Crucially, the two folded bands constructed on the reduced BZ are coupled when extended in these two additional momentum directions. 
Such a coupling is allowed because the states on the extended bands are no longer periodic in the primitive cells and thus no longer distinguished by the initial four-component momentum. 
As a result, the ensuing hybridization of the folded bands in the additional momentum directions transforms the trivial band crossings into band \emph{nodes}.
    
To understand the codimension of the resulting nodes, one can rely on the existing literature on characterizing band nodes in various symmetry classes. 
Here, we apply the results and the terminology of Ref.~\onlinecite{Bzdusek:2017}, which points out the key role of `local-in-$\vec{k}$ symmetries', i.e., those that act on Bloch Hamiltonians while leaving momentum invariant. The codimension, in turn, explains the scaling of the DOS near $E=0$.
        
In particular, the considered hyperbolic lattice model clearly has spinless time-reversal symmetry $\mathcal{T}^2=+1$ which acts through complex conjugation and maps $\mathcal{T}:\vec{k}\mapsto-\vec{k}$. 
This symmetry can potentially be combined into a local-in-$\vec{k}$ symmetry with a suitable inversion symmetry that also flips the sign of the momentum vector.
In the $2$-supercell of the $\{8,8\}$ lattice, there are several inequivalent choices for the inversion center, namely a site, the mid-point of an edge, and the center of a face.
It is derived in the Supplemental Material of Ref.~\onlinecite{Tummuru:2023}, specifically around Eq.~(F10), that it is the inversion with respect to the edge, $\mathcal{P}_\textrm{E}$, that acts on the supercell Bloch Hamiltonian by flipping the momentum sign: $\mathcal{P}_\textrm{E}:\vec{k}\mapsto - \vec{k}$. 
The composition $\mathcal{P}_\textrm{E}\mathcal{T}$ is therefore an antiunitary local-in-$\vec{k}$ symmetry with $(\mathcal{P}_\textrm{E}\mathcal{T})^2=+1$.
Additionally, the Bloch Hamiltonian exhibits a sublattice symmetry $\mathcal{S}$ because the two sites per $2$-supercell form a bipartition of the lattice.
Since the two sublattices are exchanged under inversion $\mathcal{P}_\textrm{E}$, the sublattice parity is odd~\cite{Bzdusek:2017} so that the local-in-$\vec{k}$ particle-hole symmetry squares to $(\mathcal{S}\mathcal{P}_\textrm{E}\mathcal{T})^2=-1$. 
The identified local-in-$\vec{k}$ symmetries 
\begin{equation}
    \mathcal{S},\quad(\mathcal{P}_\textrm{E}\mathcal{T})^2=+1,\quad(\mathcal{S}\mathcal{P}_\textrm{E}\mathcal{T})^2=-1
\end{equation}
locate the supercell Hamiltonian in nodal class $\textrm{CI}$, which exhibits band nodes of codimension $\delta=2$~\cite{Bzdusek:2017}.
An analogous set of local-in-$\vec{k}$ symmetries can also be identified in all the larger supercells.
Importantly, the sublattice symmetry $\mathcal{S}$ enforces a particle-hole symmetry of the spectrum at each $\vec{k}$. 
This pins all the nodes to the same energy $E=0$, resulting in the pronounced dip in the computed DOS data.

Let us next discuss the character of this DOS suppression. 
Elementary integration can be used to show that band nodes of codimension $\delta$ pinned to $E=0$ (and with the generically assumed linear dispersion) lead to the DOS scaling $\rho(E)\propto \abs{E}^{\delta-1}$~\cite{Tummuru:2023}. 
In the considered model on the $\{8,8\}$ lattice, $\delta = 2$ implies vanishing DOS at $E=0$ with linear scaling. 
This is consistent with the dip of the $2$-supercell data in \cref{main:fig:supercell:dos88}, and with the dips observed for larger supercells in \cref{fig:DOS:NN-models} above. 
    
To understand how the dip ultimately disappears in the converged DOS for large supercells, first note that the band folding at each step multiplies the number of energy bands while keeping the overall bandwidth $W$ fixed. 
Therefore, the energy extent $\Delta W$ of any individual band is expected to shrink with the size $N$ of the $N$-supercell as $\Delta W \simeq W/N$. 
(If there are $M>1$ sites per primitive cell, as is the case with $M=16$ for the $\{8,3\}$ lattice, then $\Delta W \simeq W/MN$.)
At the same time, the linear DOS scaling $\rho(E)\propto \abs{E}$ is an approximation that applies only in an energy range where the dispersion around the node can be linearly approximated. 
Since such a linear approximation cannot go beyond the energy extent of the band forming the node, the linear dip in the DOS is restricted to energies $\abs{E}\lesssim W/MN$. 
This inequality explains the rapid shrinking of the dip at $E=0$ with the growing supercell size $N$, visible in \cref{fig:DOS:NN-models:88}, and also why the dip is significantly narrower in the $\{8,3\}$ data in \cref{fig:DOS:NN-models:83}. 

We finally remark that the discussed dip is generically absent in models without sublattice/chiral symmetry. 
Since band nodes of such more general models are not pinned to a single value of energy, the node-induced DOS suppression is effectively smeared over a broader range of energies, resulting in smoother computed data.

\subsection{Nearest-neighbor models on kagome lattices}\label{Sec:kagomeNN-models}
The triangle group $\TG(2,3,p)$ allows not only constructing the $\{p,3\}$ lattice, but also its line graph, the $p$-gon-kagome lattice with sites given by $V_x\cong\rtransv{\TG^+}{\TG_x^+}$, i.e., the Schwarz triangle vertices with two-fold rotation symmetry.
Here, we consider the octagon-kagome lattice, i.e., the case $p=8$, cf.~\cref{fig:models:8kagome-NN}.
In the supplementary data and code~\cite{SDC}, we additionally provide the real-space hopping Hamiltonian in graph form as well as the raw data for the DOS.

\begin{figure}
    \centering
    \subfloat{\label{fig:DOS:others:8-kagome}}
    \subfloat{\label{fig:DOS:others:83-Haldane}}
    \includegraphics{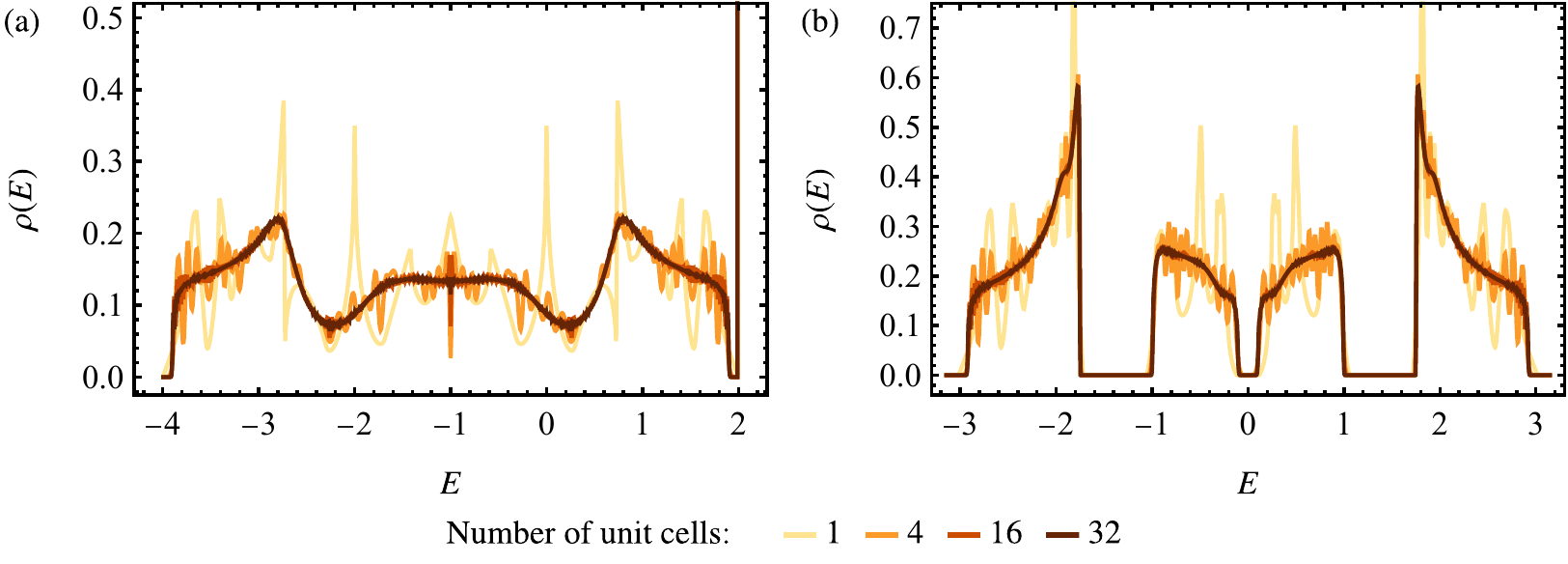}
    \caption{
        Density of states of (a) the nearest-neighbor hopping model on the octagon-kagome lattice and (b) the Haldane model on the $\{8,3\}$ lattice with $t_1=1$, $t_2=1/6$, $\phi=\pi/2$, $m=1/3$ obtained using the supercell method.
        In both cases, data on the supercell sequence $\tgquot{2}{1}$, $\tgquot{5}{1}$, $\tgquot{17}{2}$, $\tgquot{33}{1}$ with $1$, $4$, $16$, $32$ unit cells is shown and the energy resolution is $0.005$ with a moving average with window $0.01$.
    }
    \label{fig:DOS:others}
\end{figure}

As an example, we study the octagon-kagome lattice, i.e., the line graph of the $\{8,3\}$ lattice discussed above.
The supercells and the corresponding translation groups do not depend on the internal structure of the primitive cells, i.e., where orbitals are placed.
Thus, the discussion from the previous subsection applies to the octagon-kagome lattice and we can directly proceed to the results.
\Cref{main:fig:dos_data:8kagome,fig:DOS:others:8-kagome} show the DOS $\rho^{(m)}(E)$ obtained from randomly sampling $10^8$ points in $\torus{2\genus{}^{(m)}}$ for an energy resolution of $\dd{E}=0.005$ and smoothed using a moving average with window $2\dd{E}=0.01$ for the sequence in \cref{eq:supercell-sequence:238}.
Comparing \cref{fig:DOS:NN-models:83,fig:DOS:others:8-kagome}, we observe that the DOS of the NN model on the $\{8,3\}$ lattice and on its line graph match up to the flat band at $E=2$ in the latter case.
This is consistent with what is expected from graph theory~\cite{Brouwer:2012}.

\subsection{Haldane model on \texorpdfstring{$\{8,3\}$}{\{8,3\}} lattice}\label{Sec:Haldane-model}
The Haldane model on hyperbolic lattices has been introduced in Refs.~\onlinecite{Urwyler:2022,Chen:2023,Zhang:2022}.
The Hamiltonian illustrated in \cref{fig:models:83-Haldane} takes the form~\cite{Chen:2023}
\begin{equation}
    \Ham = h_1\sum_{\nn{i,j}}\left(\adjo{c_i}c_j+\text{h.c.}\right) + h_2\sum_{\overleftarrow{ij}}\left(\e^{\i\phi}\adjo{c_i}c_j+\text{h.c.}\right) + h_0\left(\sum_{i\in A}\adjo{c_i}c_i-\sum_{i\in B}\adjo{c_i}c_i\right)
\end{equation}
with NN hopping amplitude $h_1$ (solid green lines), next-NN hopping amplitude $h_2\e^{\i\phi}$ (dashed magenta lines), where $\overleftarrow{ij}$ denotes hopping from site $j$ to $i$ in mathematically positive direction (counterclockwise) for a given octagon and $\phi$ is the flux parameter.
Additionally, there is a staggered sublattice potential (called sublattice mass in the main text) $\pm h_0$ on the two sublattices $A$ and $B$ (filled and empty sites in the figure).
In the supplementary data and code~\cite{SDC}, we provide the real-space hopping Hamiltonian in graph form.
We have chosen the same parameter values as Ref.~\onlinecite{Urwyler:2022}: $h_1=1$, $h_2=1/6$, $\phi=\pi/2$, and $h_0=1/3$.

Given that the model is defined on the $\{8,3\}$ lattice, the supercells are once more the ones given in \cref{eq:supercell-sequence:238}.
In particular, the unit cell $\tgquot{2}{1}$ is exactly the primitive cell used in Ref.~\onlinecite{Urwyler:2022}, such that the DOS obtained from it matches the data presented there.
With random sampling of $10^8$ points in $\torus{2\genus{}^{(m)}}$, we have computed the DOS with an energy resolution of $\dd{E}=0.005$ and smoothed using a moving average with window $2\dd{E}=0.01$.
The resulting data is shown in \cref{main:fig:dos_data:83Haldane,fig:DOS:others:83-Haldane} and given as raw data in the supplementary data and code~\cite{SDC}. 
We observe the stability of all three energy gaps as we approach the thermodynamic limit.

\subsection{Benalcazar-Bernevig-Hughes model on \texorpdfstring{$\{6,4\}$}{\{6,4\}} lattice}\label{Sec:BBH-model}

The Benalcazar-Bernevig-Hughes (BBH) model~\cite{Benalcazar:2017}, originally defined on the square lattice, can be easily generalized to hyperbolic $\{p,4\}$ lattices with even $p$ as described in the main text for $p=6$.

\subsubsection{\texorpdfstring{$\{6,4\}$}{{6,4}} lattice}
\begin{figure}[t]
    \centering
    \includegraphics[width=0.42\linewidth]{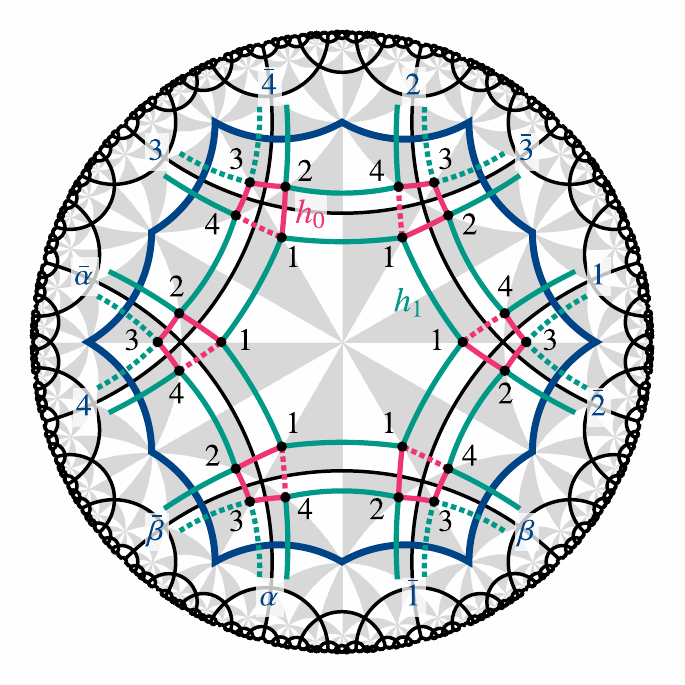}
    \caption{
        Definition of the Benalcazar-Bernevig-Hughes model on the $\{6,4\}$ lattice (black lines) and choice of gauge implementing the $\pi$-fluxes through the squares and rectangles. There are four orbitals (black dots labeled by black numbers) at each site, coupled by inter-site hoppings $\pm h_0$ (magenta) and intra-site hoppings $\pm h_1$ (green). Dashed lines indicate negative hopping amplitudes.
        The primitive cell (blue polygon) and its edge identifications are shown: the edge $\bar{1}$ is related to $1$ by the translation generator $\htgel_1$ ($\tilde{\htgel}_1$ for the supercell). Edges related by composite translations are labeled by $\alpha=\htg{3}^{-1}\htg{1}^{-1}$, $\bar{\alpha}$, $\beta=\htg{4}^{-1}\htg{2}^{-1}$, and $\bar{\beta}$.
    }
    \label{fig:bbh-model}
\end{figure}

The $\{6,4\}$ lattice is defined by the hyperbolic tessellation where four hexagons meet at each vertex,  i.e., it is formed by the four-fold rotation symmetric vertices $V_y$ of the Schwarz triangles, cf.~\cref{fig:bbh-model}.
Its space group is the triangle group $\TG(2,4,6)$ and the primitive cell (blue polygon in \cref{fig:bbh-model}) is given by the smallest quotient $\tgquot{2}{2}$ of $\TG$ with one of its normal subgroups $\HTGuc{}$~\cite{Conder:2007}:
\begin{equation}
    \HPGuc{} = \gpres{a,b,c}{a^2,b^2,c^2,(ab)^2,(bc)^4,(ca)^6,(bca^{-1}c^{-1})^2},
\end{equation}
giving
\begin{align}
    \HTGuc{} &= \gpres{\htg{1},\htg{2},\htg{3},\htg{4}}{\htg{4}\htg{1}\htg{3}\htg{4}^{-1}\htg{3}^{-1}\htg{2}^{-1}\htg{1}^{-1}\htg{2}}\\
    &= \gpres{\htg{1},\htg{2},\htg{3},\htg{4}}{\htg{1}\htg{2}\htg{3}\htg{4}\htg{3}^{-1}\htg{1}^{-1}\htg{4}^{-1}\htg{2}^{-1}}.
    \label{eq:HTG:246}
\end{align}
In Ref.~\onlinecite{Chen:2023}, a different presentation was found for this group, with six generators $\htg{1},\ldots,\htg{6}$ and three relators $\htg{1}\htg{3}\htg{5}$, $\htg{2}\htg{4}\htg{6}$, and $\htg{1}\htg{2}\htg{3}\htg{4}\htg{5}\htg{6}$. Substituting $\htg{5}=\htg{3}^{-1}\htg{1}^{-1}$ and $\htg{6}=\htg{4}^{-1}\htg{2}^{-1}$ from the first two relators into the third relator, we see that this presentation is equivalent to that found in \cref{eq:HTG:246}.

Using \gap{}, we have found a normal sequence of translation subgroups $\HTGsc{m}\nsubgstr\TG(2,4,6)$ via the appropriate quotient groups $\HPG^{(m)}\cong\TG/\HTGsc{m}$ given in Ref.~\onlinecite{Conder:2007}: With $x$, $y$, and $z$ given in \cref{eq:rotation-elements},
\begin{equation}
  \begin{alignedat}{3}
    \HPG^{(1)} &= \HPG^{\tgquot{2}{2}} &&=   \gpres{a,b,c}{a^2,b^2,c^2,x^2,y^4,z^6, (yz^{-1})^2},\\
    \HPG^{(2)} &= \HPG^{\tgquot{5}{4}} &&=   \gpres{a,b,c}{a^2,b^2,c^2,x^2,y^4,z^6, zyz^{-1}xy^{-1}z^{-2}x},\\
    \HPG^{(3)} &= \HPG^{\tgquot{9}{3}} &&=   \gpres{a,b,c}{a^2,b^2,c^2,x^2,y^4,z^6, zyxzy^2z^{-1}xzy^{-1}, zyz^{-1}yxy^{-1}z^{-1}xy^{-1}z},\\
    \HPG^{(4)} &= \HPG^{\tgquot{33}{11}} &&= \gpres{a,b,c}{a^2,b^2,c^2,x^2,y^4,z^6, xzy^{-1}z(yz^{-1})^2xy^{-1}zy^{-1}z^{-1}y},\\
    \HPG^{(4)} &= \HPG^{\tgquot{65}{9}} &&=  \gpres{a,b,c}{a^2,b^2,c^2,x^2,y^4,z^6, y^{-1}z(yz^{-1})^2yxy^{-1}z^2(y^{-1}z)^2, (xzyz^{-1})^2(xy^{-1}z^{-2})^2},
  \end{alignedat}
  \label{eq:supercell-sequence:246}
\end{equation}
with the corresponding connected symmetric supercells illustrated in \cref{fig:supercell-sequences:246} and their triangle group graphs as well as boundary identifications given in the supplementary data and code~\cite{SDC}.

\subsubsection{Benalcazar-Bernevig-Hughes model}

The BBH model is defined on the $\{6,4\}$ lattice with four orbitals on each site, i.e., on the vertices of the hexagons, cf.~\cref{fig:bbh-model}.
The four orbitals labeled by numbers $1$ to $4$ are coupled cyclically by an on-site term with amplitude $\pm h_0$, where a single one of the four couplings carries a minus sign (dashed magenta line) and the other three a plus sign (solid magenta line), thus encoding a magnetic $\pi$-flux through the cycle.
If each orbital is assigned to one of the four hexagons meeting at the lattice site, the orbitals on different sites assigned to the same hexagon are again coupled cyclically by a hopping with amplitude $\pm h_1$ (solid green line).
To encode $\pi$-fluxes through all the rectangles along the bonds of the underlying $\{6,4\}$ lattice, the green bonds forming a hexagon centered at the corner of the primitive cell carry a minus sign.
Note that while the chosen gauge does not respect all the symmetries, the physical flux pattern does, cf.~\cref{main:fig:bbh} of the main text.
This results in the following Hamiltonian
\begin{equation}
    \Ham = h_1\left[\sum_{\nn{i,j}_\mathrm{c}}\adjo{\avec{c}_i}\mqty(1&0&0&0\\0&0&0&1\\0&0&0&0\\0&0&0&0)\avec{c}_j+\sum_{\nn{i,j}_\mathrm{b}}\adjo{\avec{c}_i}\mqty(0&0&0&0\\0&0&0&1\\0&0&-1&0\\0&0&0&0)\avec{c}_j+\mathrm{h.c.}\right] + h_0\sum_i\adjo{\avec{c}_i}\mqty(0&1&0&-1\\1&0&1&0\\0&1&0&1\\-1&0&1&0)\avec{c}_i,
\end{equation}
where $\nn{i,j}_\mathrm{c}$ denotes NNs within the primitive cell, $\nn{i,j}_\mathrm{b}$ denotes NNs crossing the boundary of the primitive cell, cf.~\cref{fig:bbh-model}, and the matrices are given in the basis of orbitals $1$, $2$, $3$, $4$ (in that order).
As for the other models, we provide the real-space hopping Hamiltonian in graph form in the supplementary data and code~\cite{SDC}.

With random sampling of $10^8$ ($10^6$ for $\tgquot{33}{11}$ and $\tgquot{65}{9}$) points in $\torus{2\genus{}^{(m)}}$, we have computed the DOS with an energy resolution of $\dd{E}=0.005$ and smoothed using a moving average with window $2\dd{E}=0.01$.
The resulting data is shown in \cref{main:fig:bbh:dos} and given as raw data in the supplementary data and code~\cite{SDC}. 

\section{Convergence of the supercell method}
In this section, we provide additional data demonstrating the convergence properties of the supercell method.
First, in \cref{Sec:conv-kpoints}, we show the convergence of the density of states (DOS) obtained from a fixed supercell with increasing density of random sampling of the Abelian Brillouin zone (ABZ).
We here also observe that for the typical size of the largest supercell considered, the DOS obtained from the spectrum at $\vec{k}=0$---i.e., the spectrum of the corresponding PBC cluster---does not reproduce the characteristic features of the DOS in the thermodynamic limit; in constrast, the DOS converges rapidly upon the introduction of nontrivial momenta, i.e., under random sampling of the ABZ.
Then, in \cref{Sec:comparison-sequences}, we compare the convergence of different normal sequences of supercells, and observe that the DOS appears to converge to the same limit, thereby consolidating our expectation.

\subsection{Convergence with number of points in the Abelian Brillouin zone}\label{Sec:conv-kpoints}

Besides allowing a meaningful labeling of eigenstates and therefore simplifying a symmetry analysis~\cite{Chen:2023}, one of the main powers of the reciprocal-space description is that it allows us to work with small systems but nevertheless obtain good approximations of the spectra of larger systems.
Formally, Bloch states allow us to approximate models on large PBC clusters, i.e., \cref{eq:Ham-n-PBC:med} or even its limit for $n\to\infty$, by diagonalizing only matrices defined on smaller supercells, i.e., \cref{eq:HopMat-m-supercell,eq:Abelian-Bloch-Hamiltonian}.
To demonstrate this power, we compare 
\begin{enumerate}[(a)]
    \item the DOS obtained from treating a given supercell as a PBC cluster in the form of \cref{eq:Ham-m-PBC} with matrix $\HopMat^{uv}_{ij}\left(\htgel\right)=h^{uv}\left(\eta_i\htgel\eta_j^{-1}\right)$, which corresponds to evaluating the Abelian Bloch Hamiltonian in \cref{eq:Abelian-Bloch-Hamiltonian} at $\vec{k}=0$, to 
     \item the DOS obtained from treating the same supercell as a unit cell of a larger PBC cluster, where the same matrix $\HopMat^{uv}_{ij}\left(\htgel\right)$ appears in \cref{eq:Ham-n-PBC-full:Bloch-decomp}, and randomly sampling the ABZ given in \cref{eq:ABZ} with increasing number of sampled points.
\end{enumerate}
Note that (a) is, in essence, the approach of Ref.~\onlinecite{Lux:2023}, where the authors diagonalize $\HopMat^{uv}_{ij}$ on extremely large PBC clusters whose associated translation groups also form sequences of normal subgroups, albeit different ones from the ones studied here.

\begin{figure}
    \centering
    \subfloat{\label{fig:DOS:k-conv:88:PBC}}
    \subfloat{\label{fig:DOS:k-conv:88:1DIRs}}
    \includegraphics{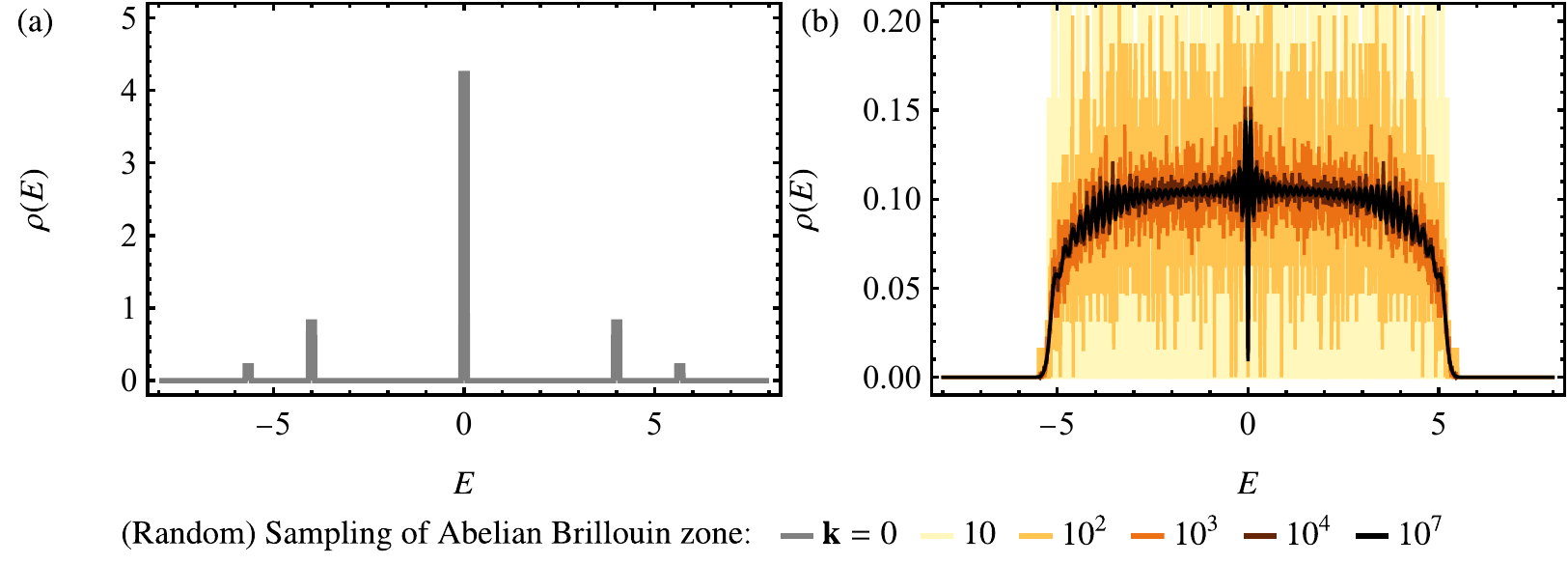}
    \caption{
        Convergence of density of states (DOS) of the nearest-neighbor hopping model defined on the $64$-supercell $\tgquot{65}{78}$ of the $\{8,8\}$ lattice.
        (a) DOS obtained from treating the $64$-supercell as a PBC cluster with $64$ sites, effectively setting the Abelian momentum $\vec{k}=0$. The data has been obtained with a resolution of $\dd{E}=0.005$ and smoothed with a moving average over windows of width $0.08$. 
        (b) DOS obtained from treating the $64$-supercell as the unit cell of a large PBC cluster and randomly sampling Abelian momenta. The number of sampled momenta is given in the legend. The data has been obtained with a resolution of $\dd{E}=0.005$ and smoothed with a moving average over windows of width $0.01$.
    }
    \label{fig:DOS:k-conv:88}
\end{figure}

\begin{figure}
    \centering
    \subfloat{\label{fig:DOS:k-conv:83:PBC}}
    \subfloat{\label{fig:DOS:k-conv:83:1DIRs}}
    \includegraphics{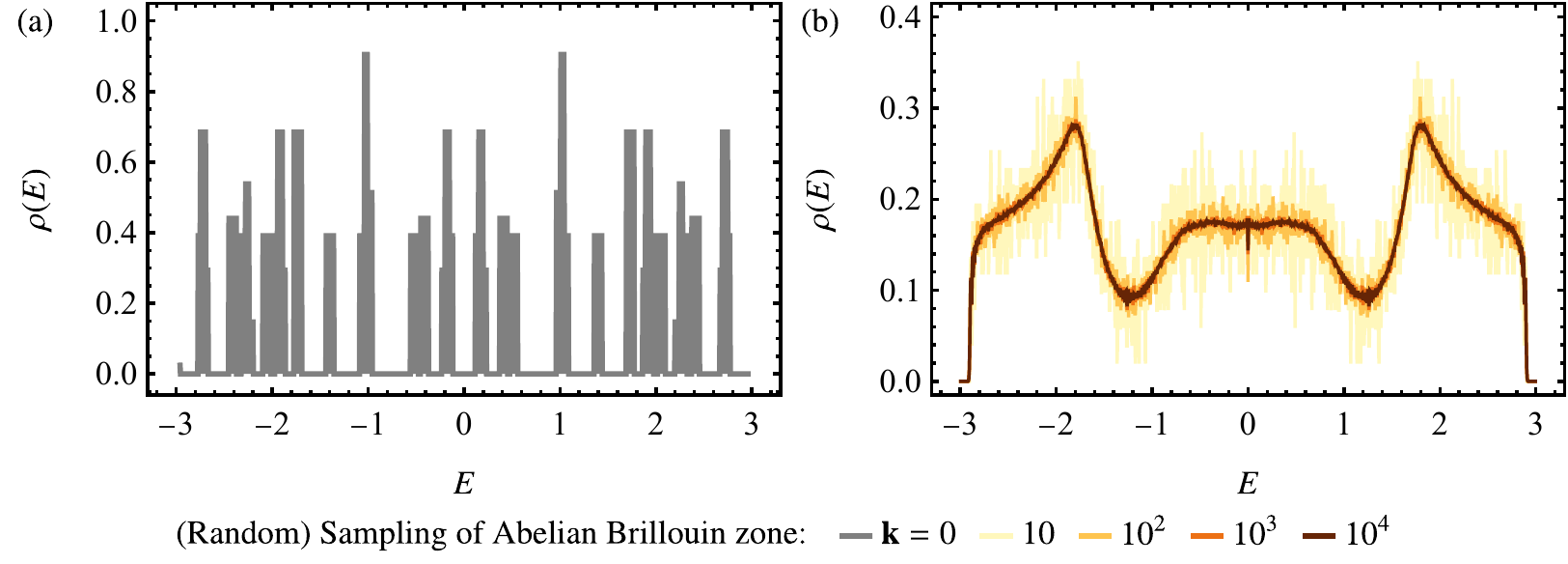}
    \caption{
        Convergence of density of states (DOS) of the nearest-neighbor hopping model defined on the $32$-supercell $\tgquot{33}{1}$ of the $\{8,3\}$ lattice.
        (a) DOS obtained from treating the $32$-supercell as a PBC cluster with $512$ sites, effectively setting the Abelian momentum $\vec{k}=0$. The data has been obtained with a resolution of $\dd{E}=0.005$ and smoothed with a moving average over windows of width $0.16$.
        (b) DOS obtained from treating the $32$-supercell as the unit cell of a large PBC cluster and randomly sampling Abelian momenta. The number of sampled momenta is given in the legend. The data has been obtained with a resolution of $\dd{E}=0.005$ and smoothed with a moving average over windows of width~$0.01$.
    }
    \label{fig:DOS:k-conv:83}
\end{figure}

In \cref{fig:DOS:k-conv:88}, we show the data described above for the NN model on the supercell $\tgquot{65}{78}$ of the $\{8,8\}$ lattice.
The DOS obtained from simply diagonalizing the $64$-site PBC cluster is shown in \cref{fig:DOS:k-conv:88:PBC} and is very far from the converged DOS in \cref{fig:DOS:NN-models:88}.
On the other hand, \cref{fig:DOS:k-conv:88:1DIRs} shows that upon introducing Abelian momenta, the DOS rapidly converges, with $10^4$ (dark red line) randomly sampled points already reproducing most of the features, and $10^7$ (black line) being basically indistinguishable from the converged curve shown in \cref{fig:DOS:NN-models:88}.
Note that here we are considering convergence for a \emph{fixed} supercell in the number of sampled momenta, and not in the sequence of supercells; in the latter case $\tgquot{65}{78}$ would be considered not fully converged.

We repeat the same analysis for the NN model on the smaller supercell $\tgquot{33}{1}$ of the $\{8,3\}$ lattice, which contains $32\times 16 = 512$ sites, cf.~\cref{fig:DOS:k-conv:83}.
As for the previous model, the DOS obtained from simply diagonalizing the $512$-site PBC cluster is shown in \cref{fig:DOS:k-conv:83:PBC} and is very far from the converged DOS in \cref{fig:DOS:NN-models:83}.
On the other hand, \cref{fig:DOS:k-conv:83:1DIRs} shows again that upon introducing Abelian momenta, the DOS rapidly converges with $100$ (light orange line) randomly sampled points already reproducing most of the features and $10^4$ (dark red line) being basically indistinguishable from the converged curve shown in \cref{fig:DOS:NN-models:83}.
A similar convergence can be observed for the Haldane model on the same supercell.

\subsection{Comparison of different supercell sequences}\label{Sec:comparison-sequences}
\begin{figure}
    \centering
    \subfloat{\label{fig:DOS:sequences:88:2}}
    \subfloat{\label{fig:DOS:sequences:88:3}}
    \includegraphics{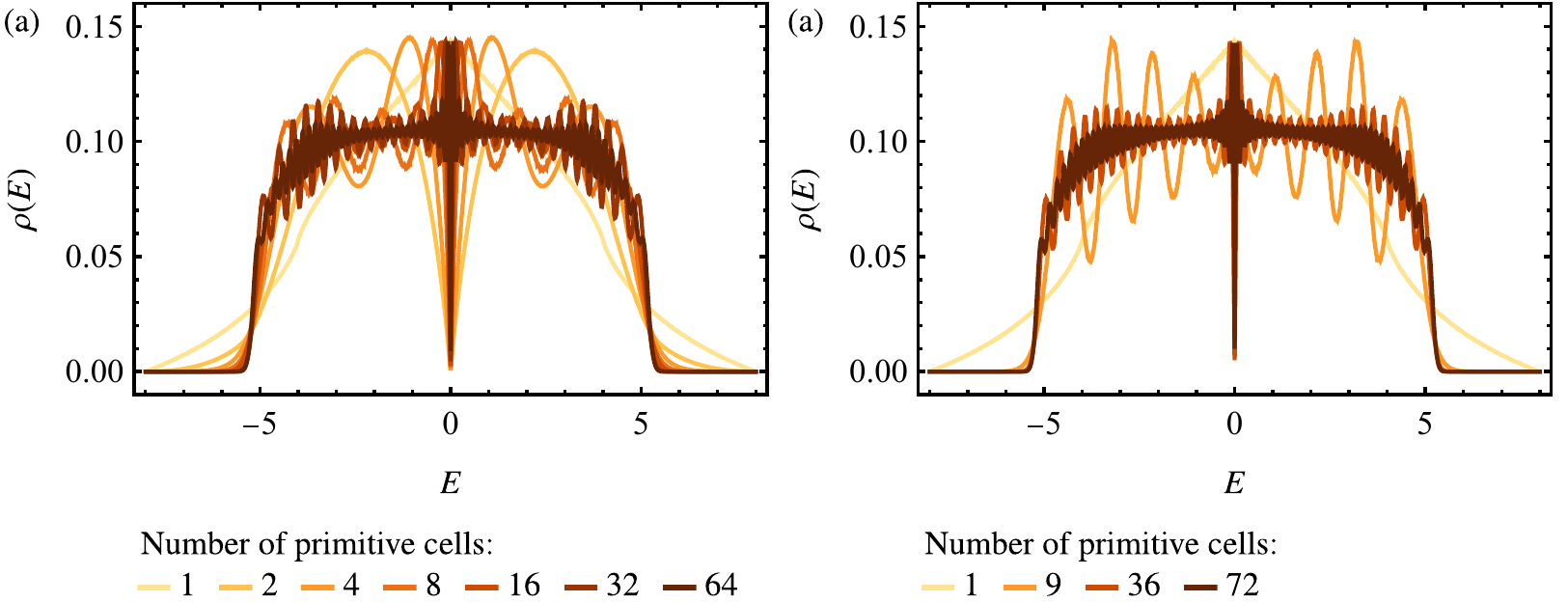}
    \caption{
        Density of states of the nearest-neighbor hopping model on the $\{8,8\}$ lattice [cf.~\cref{fig:DOS:NN-models:88}] for (a) the supercell sequence $\tgquot{2}{6}$, $\tgquot{3}{11}$, $\tgquot{5}{13}$, $\tgquot{9}{22}$, $\tgquot{17}{35}$, $\tgquot{33}{58}$, $\tgquot{65}{81}$ with $1$, $2$, $4$, $8$, $16$, $32$, $64$ unit cells per supercell and (b) the sequence $\tgquot{2}{6}$, $\tgquot{10}{22}$, $\tgquot{37}{37}$, $\tgquot{73}{71}$ with $1$, $9$, $36$, $72$ unit cells per supercell.
        In both cases, the energy resolution is $0.005$ with a moving average with window $0.01$.
    }
    \label{fig:DOS:sequences:88}
\end{figure}

\begin{figure}
    \centering
    \subfloat{\label{fig:DOS:sequences:83:2}}
    \subfloat{\label{fig:DOS:sequences:83:3}}
    \includegraphics{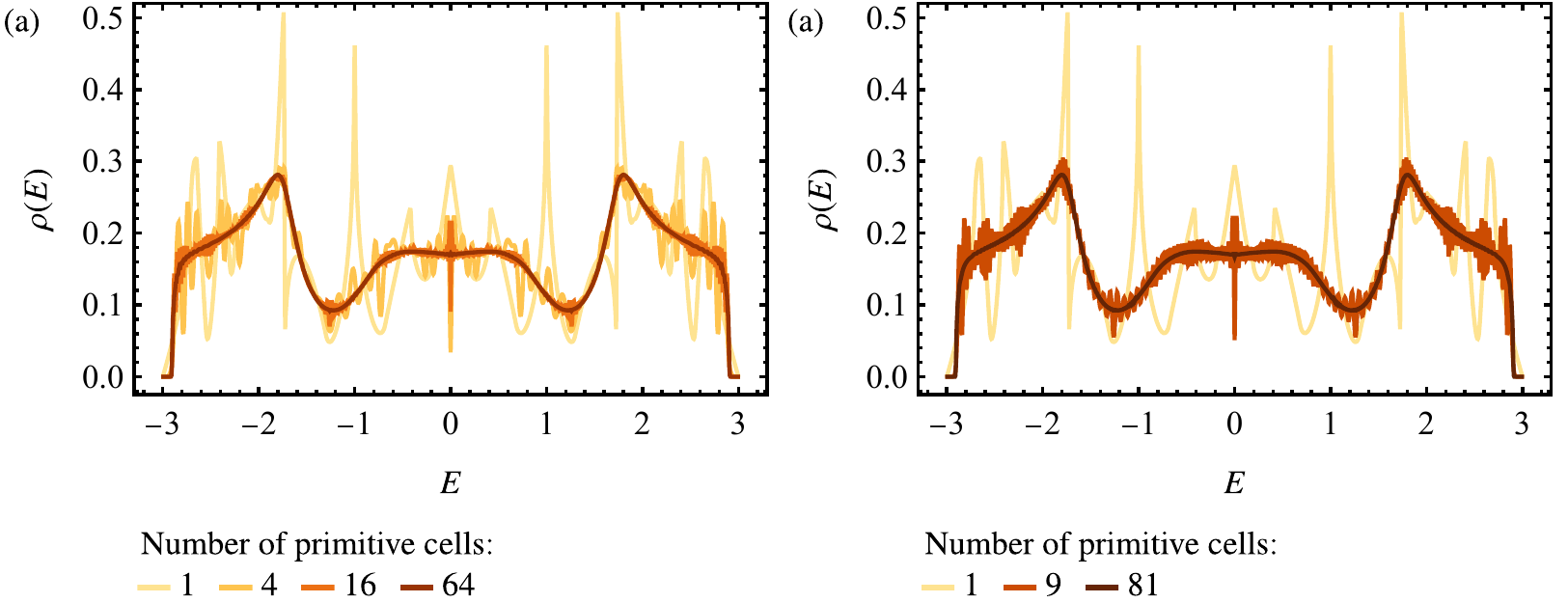}
    \caption{
        Density of states of the nearest-neighbor hopping model on the $\{8,3\}$ lattice [cf.~\cref{fig:DOS:NN-models:83}] for (a) the supercell sequence $\tgquot{2}{1}$, $\tgquot{5}{1}$, $\tgquot{17}{2}$, $\tgquot{65}{1}$ with $1$, $4$, $16$, $64$ unit cells per supercell and (b) the sequence $\tgquot{2}{1}$, $\tgquot{10}{1}$, $\tgquot{28}{1}$, $\tgquot{82}{1}$ with $1$, $9$, $27$, $81$ unit cells per supercell.
        In both cases, the energy resolution is $0.005$ with a moving average with window $0.01$.
    }
    \label{fig:DOS:sequences:83}
\end{figure}

Since the supercell method does not produce unique normal sequences but allows for many different sequences for the same lattice and choice of primitive cell, it remains to be shown that different sequences converge to the same limit.
We demonstrate this here using again the NN hopping models on the $\{8,8\}$ and the $\{8,3\}$ lattice.
For the $\{8,8\}$ lattice, alternative normal sequences to the one given in \cref{eq:supercell-sequence:288} are~\cite{Conder:2007}
\begin{gather}
    \tgquot{2}{6}, \tgquot{3}{11}, \tgquot{5}{13}, \tgquot{9}{22}, \tgquot{17}{35}, \tgquot{33}{58}, \tgquot{65}{81};\\
    \tgquot{2}{6}, \tgquot{10}{22}, \tgquot{37}{37}, \tgquot{73}{71},
\end{gather}
and the corresponding DOS data for $10^7$ randomly sampled points in $\torus{2\genus{}^{(m)}}$ is shown in \cref{fig:DOS:sequences:88:2,fig:DOS:sequences:88:3}, respectively.
Comparing those two figures to \cref{fig:DOS:NN-models:88}, we observe that they apparently converge to the same limit.
For the $\{8,3\}$ lattice, alternative normal sequences to the one given in \cref{eq:supercell-sequence:238} are~\cite{Conder:2007}
\begin{gather}
    \tgquot{2}{1}, \tgquot{5}{1}, \tgquot{17}{2}, \tgquot{65}{1};\\
    \tgquot{2}{1}, \tgquot{10}{1}, \tgquot{28}{1}, \tgquot{82}{1},
\end{gather}
and the corresponding DOS data for $10^7$ randomly sampled points in $\torus{2\genus{}^{(m)}}$ is shown in \cref{fig:DOS:sequences:83:2,fig:DOS:sequences:83:3}, respectively.
Comparing those two figures to \cref{fig:DOS:NN-models:83}, we again observe apparent convergence to the same limit.

\subsection{Comparison to continued-fraction method}\label{Sec:comparison-cfmethod}

\begin{figure}
    \centering
    \subfloat{\label{fig:DOS:comp-cfm:88}}
    \subfloat{\label{fig:DOS:comp-cfm:83}}
    \includegraphics{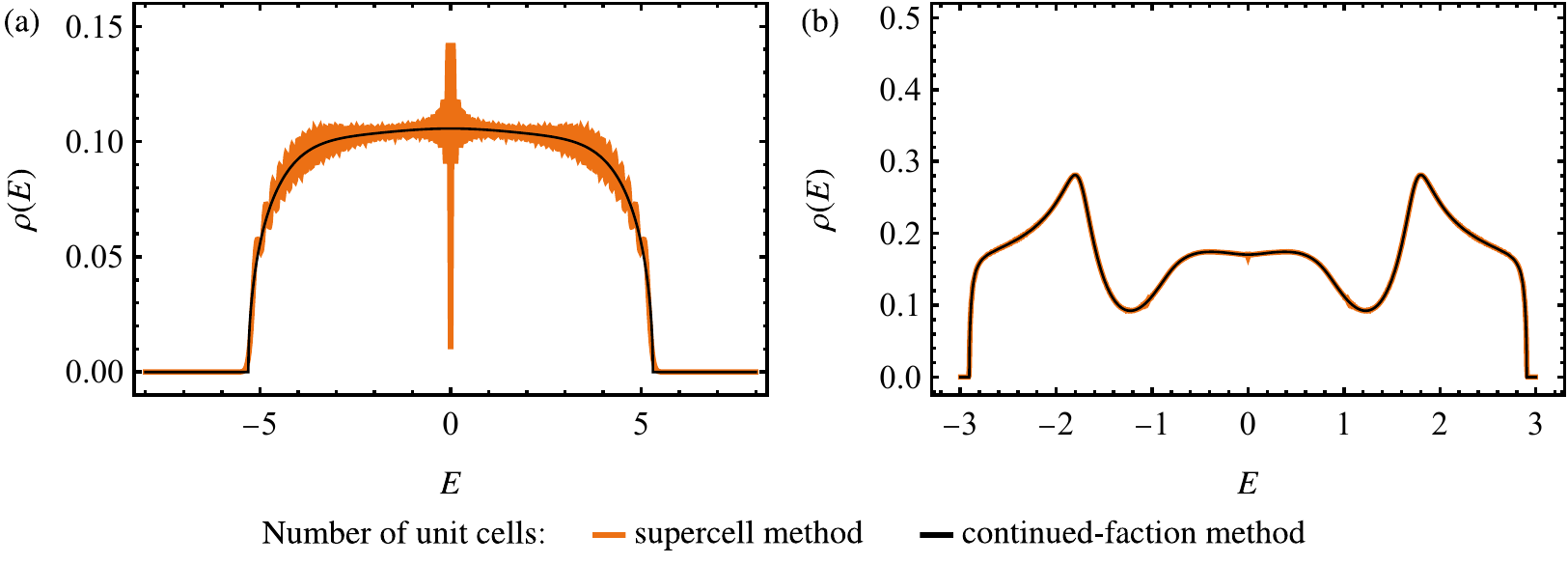}
    \caption{
        Density of states of the nearest-neighbor hopping model on (a) the $\{8,8\}$ lattice and (b) the $\{8,3\}$ lattice obtained using the supercell method (orange) and the continued-fraction method (black; results from Ref.~\onlinecite{Mosseri:2023}).
        In (a) the supercell $\tgquot{73}{71}$ with $72$ sites and in (b) $\tgquot{82}{1}$ with $1296$ sites was used and in both cases, the energy resolution is $0.005$ with a moving average with window $0.01$.
    }
    \label{fig:DOS:comp-cfm}
\end{figure}

The DOS of the very simple gapless nearest-neighbor hopping models was studied using the continued-fraction method in Ref.~\onlinecite{Mosseri:2023}.
We here compare our results obtained for supercells of the $\{8,8\}$ and $\{8,3\}$ lattice that contain $72$ and $1296$ sites, respectively, to the results of Ref.~\onlinecite{Mosseri:2023} obtained from more than $10^9$ sites.
\Cref{fig:DOS:comp-cfm} compares the DOS obtained from the two methods and we observe that in both cases the data obtained from the supercell method clearly converges to those obtained using the continued-fraction method.
In particular for the $\{8,3\}$ lattice (\cref{fig:DOS:comp-cfm:83}), almost no difference is discernible.
On the other hand, our results for the $\{8,8\}$ lattice are clearly not fully converged (which can be attributed to the much lower number of sites in the corresponding supercell), but the core features are reproduced.

For a more quantitative comparison of the data, we compute the moments of the density of states.
The $n^\mathrm{th}$ moment $M_n$ of $\rho(E)$ is defined as
\begin{equation}
    M_n = \int_{-\infty}^{\infty}\dd{E}\rho(E)E^n.
\end{equation}
Due to the symmetry of the spectra for both the $\{8,8\}$ as well as the $\{8,3\}$ lattice, all odd moments vanish.
In \cref{tab:dos-moments:88,tab:dos-moments:83}, we compare the moments obtained from the supercells discussed above to those obtained analytically in Ref.~\onlinecite{Mosseri:2023}.
For $\{8,8\}$, we observe that although the density of states obtained from the $72$-site supercell $\tgquot{73}{71}$ is clearly not yet converged (cf.~\cref{fig:DOS:comp-cfm:88}), the first three even moments are reproduced exactly and the relative error remains below $10^{-4}$ up to the $16^\mathrm{th}$ moment.
The accuracy of the odd moments, which should all vanish identically, is similar: the first five odd moments are below $10^{-10}$ and up to the $15^\mathrm{th}$ moment, all of them are below $10^{-4}$.
For $\{8,3\}$, \cref{fig:DOS:comp-cfm:83} indicates a significantly better level of convergence and we find accordingly that the first seven even moments, i.e., up to the $14^\mathrm{th}$ moment, are obtained exactly, while the relative error remains below $10^{-4}$ for the first 18 even moments.
The first seven odd moments are below $10^{-10}$ and up to the $27^\mathrm{th}$ moment, all of them are below $10^{-4}$.

\begin{table}
    \begin{ruledtabular}
    \centering
    \begin{tabular}{rS[table-format=1.2e2]S[table-format=1]crrS[table-format=16]}
        \multicolumn{3}{c}{Odd moments} & & \multicolumn{3}{c}{Even moments}\\\cline{1-3}\cline{5-7}
        $n$ & \multicolumn{1}{l}{Supercell} & \multicolumn{1}{l}{Continued-fraction~\cite{Mosseri:2023}} & \rule{0pt}{1.1em} & $n$ & \multicolumn{1}{l}{Supercell} & \multicolumn{1}{l}{Continued-fraction~\cite{Mosseri:2023}} \\[0.1em]\cline{1-3}\cline{5-7}\addlinespace
         1  & 1.55e-16 & 0 & \hfill & 2  & {8}                & 8                \\
         3  & 8.92e-15 & 0 & \hfill & 4  & {120}              & 120              \\
         5  & 2.66e-13 & 0 & \hfill & 6  & {2\,192}          & 2192             \\
         7  & 7.48e-12 & 0 & \hfill & 8  & {44\,26\textcolor{orange}{3}}          & 44264            \\
         9  & 2.02e-10 & 0 & \hfill & 10 & {950\,\textcolor{orange}{582}}           & 950608           \\
         11 & 5.19e-9  & 0 & \hfill & 12 & {21\,288\,\textcolor{orange}{185}}         & 21288912         \\
         13 & 1.45e-7  & 0 & \hfill & 14 & {491\,5\textcolor{orange}{03\,907}}        & 491515088        \\
         15 & 3.96e-6  & 0 & \hfill & 16 & {11\,614\,\textcolor{orange}{603\,543}}      & 11614244072      \\
         17 & 1.10e-4  & 0 & \hfill & 18 & {279\,\textcolor{orange}{540\,473\,553}}     & 279495834368     \\
         19 & 2.97e-3  & 0 & \hfill & 20 & {6\,82\textcolor{orange}{8\,817\,799\,853}}    & 6826071585040    \\
         21 & 8.18e-2  & 0 & \hfill & 22 & {168\,\textcolor{orange}{891\,218\,655\,925}}  & 168755930104880  \\
         23 & 2.24     & 0 & \hfill & 24 & {4\,2\textcolor{orange}{20\,862\,714\,087\,351}} & 4214946994935248 \\
    \end{tabular}
    \end{ruledtabular}
    \caption{
        Odd (left) and even (right) moments of the density of states of the $\{8,8\}$ lattice obtained from the supercell (based on $\tgquot{73}{71}$ and rounded to the nearest integer) and continued-fraction~\cite{Mosseri:2023} methods. The numbers given by the latter are exact. 
        For even moments, digits highlighted in orange indicate deviation from the exact result.
    }
    \label{tab:dos-moments:88}
\end{table}

\begin{table}
    \begin{ruledtabular}
    \centering
    \begin{tabular}{rS[table-format=1.2e2]S[table-format=1]crrS[table-format=16]}
        \multicolumn{3}{c}{Odd moments} & & \multicolumn{3}{c}{Even moments}\\\cline{1-3}\cline{5-7}
        $n$ & \multicolumn{1}{l}{Supercell} & \multicolumn{1}{l}{Continued-fraction~\cite{Mosseri:2023}} & \rule{0pt}{1.1em} & $n$ & \multicolumn{1}{l}{Supercell} & \multicolumn{1}{l}{Continued-fraction~\cite{Mosseri:2023}} \\[0.1em]\cline{1-3}\cline{5-7}\addlinespace
          1 & 9.13e-17 & 0 & \hfill &   2 & {3}               & 3\\
          3 & 6.37e-16 & 0 & \hfill &   4 & {15}              & 15\\
          5 & 3.59e-15 & 0 & \hfill &   6 & {87}              & 87\\
          7 & 4.09e-14 & 0 & \hfill &   8 & {549}             & 549\\
          9 & 3.06e-13 & 0 & \hfill &  10 & {3\,663}            & 3663\\
         11 & 2.76e-12 & 0 & \hfill &  12 & {25\,407}           & 25407\\
         13 & 1.80e-11 & 0 & \hfill &  14 & {181\,233}          & 181233\\
         15 & 1.93e-10 & 0 & \hfill &  16 & {1\,320\,11\textcolor{orange}{9}}         & 1320117\\
         17 & 1.47e-9  & 0 & \hfill &  18 & {9\,772\,3\textcolor{orange}{95}}         & 9772359\\
         19 & 1.05e-8  & 0 & \hfill &  20 & {73\,27\textcolor{orange}{4\,242}}        & 73273755\\
         21 & 8.61e-8  & 0 & \hfill &  22 & {555\,1\textcolor{orange}{64\,111}}       & 555158277\\
         23 & 8.27e-7  & 0 & \hfill &  24 & {4\,242\,6\textcolor{orange}{68\,085}}      & 4242602877\\
         25 & 5.30e-6  & 0 & \hfill &  26 & {32\,6\textcolor{orange}{60\,090\,468}}     & 32659394745\\
         27 & 4.82e-5  & 0 & \hfill &  28 & {252\,98\textcolor{orange}{7\,887\,054}}    & 252980710305\\
         29 & 4.28e-4  & 0 & \hfill &  30 & {1\,970\,\textcolor{orange}{260\,645\,697}}   & 1970188493067\\
         31 & 3.80e-3  & 0 & \hfill &  32 & {15\,416\,\textcolor{orange}{884\,219\,376}}  & 15416173400134\\
         33 & 2.94e-2  & 0 & \hfill &  34 & {121\,13\textcolor{orange}{7\,511\,606\,951}} & 121130623234816\\
         35 & 2.21e-1  & 0 & \hfill &  36 & {955\,3\textcolor{orange}{67\,811\,569\,773}} & 955301961767219\\
    \end{tabular}
    \end{ruledtabular}
    \caption{
        Odd (left) and even (right) moments of the density of states of the $\{8,3\}$ lattice obtained from the supercell (based on $\tgquot{82}{1}$ and rounded to the nearest integer) and continued-fraction~\cite{Mosseri:2023} methods. The numbers given by the latter are exact.
        For even moments, digits highlighted in orange indicate deviation from the exact result.
    }
    \label{tab:dos-moments:83}
\end{table}

Overall, the comparison with the exact moments gives further evidence to the convergence of the supercell method and affirms our expectation that other quantities besides the density of states, where our reciprocal-space perspective can fully manifest its efficacy, will experience similar convergence.


\end{document}